\documentclass[reprint,twocolumn,groupedaddress,showkeys,floatfix,superscriptaddress,aps]{revtex4-2}
\usepackage[colorlinks=true,citecolor=red,filecolor=green,linkcolor=blue,pdfnewwindow=true]{hyperref}
\usepackage{amsmath} 
\usepackage{amssymb}
\usepackage[version=4]{mhchem}
\usepackage{graphicx}
\usepackage{bm}
\usepackage[title]{appendix} 
\usepackage{float}
\usepackage[normalem]{ulem}
\usepackage[dvipsnames]{xcolor}
\usepackage{setspace}
\usepackage{algorithm} 
\usepackage{algpseudocode}

\hypersetup{urlcolor=blue}

\makeatletter
 
\newcommand{\Rmnum}[1]{\expandafter\@slowromancap\romannumeral #1@}

\makeatother

\begin{document}
\title{Quantifying dissipation in stochastic complex oscillations}
\author{Athokpam Langlen Chanu}   \email{athokpam.chanu@apctp.org (Corresponding author)}    
\affiliation{Asia Pacific Center for Theoretical Physics, Pohang, 37673, Republic of Korea}

\author{Preet Mishra}\email{preet45$_$sit@jnu.ac.in} 
\affiliation{School of Computational and Integrative Sciences, Jawaharlal Nehru~University, New Delhi, 110067, India}  

\author{Shyam Kumar} \email{shyam49$_$sit@jnu.ac.in}  
\affiliation{School of Computational and Integrative Sciences, Jawaharlal Nehru~University, New Delhi, 110067, India}

\author{R.~K.~Brojen Singh} \email{brojen@jnu.ac.in}  
\affiliation{School of Computational and Integrative Sciences, Jawaharlal Nehru~University, New Delhi, 110067, India} 
\begin{abstract}
Fluctuations-driven complex oscillations are experimentally observed in cellular systems such as hepatocytes, cardiac cells, neuronal cells, etc.  These systems are generally operating in regimes far from thermodynamic equilibrium. To study nonequilibrium thermodynamic properties such as energy dissipation in stochastic complex oscillations, we consider stochastic modeling of two nonlinear biological oscillators, namely, the intracellular calcium (Ca$^{2+}$) oscillation model and the Hindmarsh-Rose model for neuronal dynamics. These models exhibit various types of complex oscillations such as bursting and quasi-periodic oscillations for various system parameter values. In this work, we formulate open chemical reaction schemes of the two model systems driving the systems far from thermodynamic equilibrium. We then analyze the steady-state total entropy production rate ($\dot{S}^{\mathrm{ss}}_{\mathrm{tot}}$) in the various types of stochastic complex oscillations. Our results show higher values of $\dot{S}^{\mathrm{ss}}_{\mathrm{tot}}$ in stochastic complex oscillations than simple periodic oscillations. Moreover, in the Hindmarsh-Rose neuronal model, we observe an order-to-disorder transition from periodic (organized) bursts of spikes to chaotic (unorganized) oscillations with distinct behaviors of $\dot{S}^{\mathrm{ss}}_{\mathrm{tot}}$.Our results reveal that stochastic complex oscillations are produced at the cost of higher energy consumption and that it requires a higher thermodynamic cost to maintain the periodic bursts than chaotic oscillations. Our findings indicate that complex cellular regulatory or signaling processes by Ca$^{2+}$ that help in performing complex tasks of the nervous system or  rich information coding by neurons involve a higher thermodynamic cost. The results obtained deepen our understanding of energy dissipation in nonlinear, nonequilibrium biological systems with stochastic complex oscillatory dynamics. 


\end{abstract}

\maketitle
\section{Introduction}
Complex systems are characterized by the emergent properties arising out of the interplay of numerous sub-units or components within them, engaging in complicated mutual interactions~\cite{ladyman2013complex}. For instance, biological living cells are complex biochemical systems where a multitude of interactions occur among a number of chemical or molecular species at various levels of organization~\cite{ma2017complex}. These interactions can be described by a general network of chemical reactions, referred to as a chemical reaction network (CRN). If $\{X_i\}$; $i=1,2,...,k$ represents a set of chemical species undergoing $M$ chemical reactions, then CRN is given by
\small{\begin{equation}
\ce{\sigma_{1m}X_1 + \sigma_{2m}X_2 + ... + \sigma_{km}X_k ->[\eta_m] \rho_{1m}X_1 + \rho_{2m}X_2 + ... + \rho_{km}X_k},\label{crn1}
\end{equation}}
\normalsize 
{\noindent}where $\eta_m$ represents the reaction rate constant of reaction $m$ $(=1,2,\dots,M)$. Denoting $\{x_1,x_2,\dots\}$ as the corresponding concentrations of $\{X_1,X_2,\dots\}$ species, according to the law of mass action, the reaction rate equations governing the time evolution of the concentrations are given by
\small{
\begin{align}
\dot{x_1}&=f_1(x_1,x_2,\dots,x_k,\Lambda),\nonumber \\
&\vdots \label{eq:rbnew22} \\
\dot{x_k}&=f_k(x_1,x_2,\dots,x_k,\Lambda).\nonumber
\end{align}
}
\normalsize
{\noindent}The functions $\{f_i\}$ in the coupled ordinary differential equations (ODEs)~\eqref{eq:rbnew22} represent the complex interactions among the species. The parameter $\Lambda$ denotes the control parameter, whose changes can lead to the nonlinear phenomenon of bifurcation, resulting in diverse nonlinear chemical kinetics~\cite{strogatz2018nonlinear}. The steady-state concentrations $\{x^{*}_1,x^{*}_2,\dots\}$ can be obtained by setting and solving $\dot{x_1}=\dot{x_2}=\dots=0$. Within the framework of linear stability analysis~\cite{strogatz2018nonlinear}, the stability of these states can be assessed by examining the eigenvalues $\lambda$ obtained from the characteristic equation. A positive real part of eigenvalues (i.e., Re$(\lambda_{\pm}) > 0$) indicates unstable steady-states, leading to Hopf bifurcation or limit cycle~\cite{strogatz2018nonlinear}. For example, in biochemical systems, changes in the control parameter $\Lambda$ can induce Hopf bifurcation, resulting in biochemical oscillation~\cite{cao2022improved}. Consequently, CRN~\eqref{crn1} can exhibit diverse dynamical behaviors, including steady-states (fixed points), bistability, tristability, excitability, and oscillations (limit cycles). Building upon seminal theoretical works of Turing on spontaneous pattern formation in reaction-diffusion systems~\cite{turing1990chemical}, Prigogine and co-workers explored nonlinear chemical kinetics and developed the theory of dissipative structures in nonlinear, nonequilibrium thermodynamics~\cite{glansdorff1971thermodynamic, prigogine1978time,kondepudi2014modern}. Chemical instability induces bifurcation~\cite{strogatz2018nonlinear} that leads to dissipative structures, fostering the nonlinear phenomenon of self-organization~\cite{prigogine1977self,epstein2006introduction, yates2012self,chung2022thermodynamics}. Chemical reaction systems represented by CRN~\eqref{crn1} can exhibit various types of dissipative structures, including temporal oscillations, symmetry-breaking instabilities, and multiple kinetic steady-states resulting in hysteresis effects~\cite{noyes1974oscillatory}. Investigating the nonequilibrium thermodynamic properties associated with diverse dissipative structures is not only of fundamental significance but also holds practical applications in fields such as biochemical sciences, materials science, and chemical engineering. 

Oscillatory dynamics, a prevalent nonlinear phenomenon, manifests in various natural systems, spanning chemical, biological, and physical domains. The phenomenon of oscillation is a well-established nonequilibrium phenomenon~\cite{prigogine1977self}. Biological rhythms~\cite{goldbeter2008biological}, such as glycolytic oscillations~\cite{albert1997biochemical}, circadian rhythms~\cite{albert1997biochemical,goldbeter2007biological}, cell cycles~\cite{ferrell2011modeling}, calcium signaling~\cite{berridge1994spatial}, and cyclic AMP signaling~\cite{martiel1985autonomous} are recognized dissipative structures~\cite{goldbeter2018dissipative}. These biological processes generally operate by consuming free energy (e.g., ATP or GTP) in the system and dissipating energy into the environment~\cite{cao2015free,song2021thermodynamic,ghosh2023developmental}. Such processes in microscopic and mesoscopic biological systems display fluctuation-driven dynamics, as experimentally observed~\cite{tsimring2014noise}. These fluctuations in dissipative, nonequilibrium systems have been of paramount interest to theoreticians and experimentalists. Estimating energy dissipation, quantified by entropy production, in nonequilibrium systems driven by fluctuations is an active area of research~\cite{van2020entropy,skinner2021improved,yu2021inverse,manikandan2021quantitative,otsubo2022estimating,foster2023dissipation}. For reviews on energy dissipation and inference in living systems, see e.g.,~\cite{fang2020nonequilibrium,lucia2020time,kim2023free}. Some recent studies that have analyzed dissipation in stochastic systems exhibiting oscillatory dynamics include~\cite{rana2020precision,oberreiter2022universal,cao2022improved,cao2023cascade}. However, while these recent studies predominantly focus on oscillations with a single periodicity (termed as simple periodic oscillation), complex oscillatory behaviors - including bursts of spikes with multiple periodicities, bursting dynamics of a single sharp spike followed by smaller amplitude spikes about a plateau region, or quasi-periodic oscillations with two or more frequencies - are observed experimentally in various biological cells, such as hepatocytes, cardiac cells, neurons and human embryonic kidney cells~\cite{borghans1997complex,terrar2020calcium,skinner2021improved}. In nonlinear dynamics, deterministic complex oscillations are often analyzed using methods such as Jacobian and Lyapunov exponents, power spectra, or return maps~\cite{houart1999bursting}. However, since the complex oscillations observed in real-world biological systems (examples mentioned above) are fluctuation-driven processes, one needs to address the properties of these systems within a stochastic framework. 
In such situations, energy dissipation in stochastic complex oscillations, encompassing bursting, and quasi-periodicity, remains poorly understood. The present paper aims to bridge this research gap by investigating dissipation in two well-known nonlinear models of biological dynamical oscillators with complex oscillations: intracellular calcium (Ca$^{2+}$) oscillation model and Hindmarsh-Rose (HR) neuronal model. 

Calcium ions (Ca$^{2+}$) serve as crucial messengers within biological cells. 
Intracellular Ca$^{2+}$ oscillations, pivotal in processes like cell signaling, muscle contraction, gene expression, and cell differentiation~\cite{skinner2021improved}, are observed experimentally in various cell types, including hepatocytes~\cite{wu2005phase}, pancreatic cells~\cite{tamarina2005inositol}, muscle cells~\cite{collier2000calcium, meng2007calcium}, and neuronal cells~\cite{verkhratsky1996calcium}. These oscillations exhibit various complex phenomena, featuring bursting, period doubling, chaos, and quasi-periodicity which could correspond to various cellular functions~\cite{houart1999bursting}. These complex oscillations are 
far from thermodynamic equilibrium phenomena driven by several factors.
Nonequilibrium dynamics and thermodynamics of intracellular Ca$^{2+}$ oscillations were explored in Ref.~\cite{xu2012potential} using potential landscape and flux theory. However, this study is limited to simple periodic oscillation. Again, several studies on the well-known Hindmarsh-Rose (HR) neuronal model were performed primarily by focusing on the information theory of neuron dynamics~\cite{savin2017maximum,shimazaki2015neurons}. However, recent studies have begun to consider the thermodynamic aspects of neuronal cells.
These include investigations into the thermodynamics of single neurons~\cite{wang2023global}, thermodynamic constraints on neural dimensions and firing rates~\cite{karbowski2009thermodynamic}, thermodynamics of learning in single neurons~\cite{goldt2017stochastic}, thermodynamic costs of information processing~\cite{sartori2014thermodynamic}, and thermodynamics of action potential propagation~\cite{andersen2009towards,drukarch2022thermodynamic,wang2017studies}. Further, energy analysis of bursting in neurons has been discussed in~\cite{moujahid2022energy}, while fluctuation-dissipation relations for neuronal dynamics have been reported recently~\cite{lindner2022fluctuation}. In this paper, we adopt a new approach to the intracellular calcium oscillation model and the Hindmarsh-Rose neuronal model by formulating them as stochastic nonlinear chemical oscillators far from thermodynamic equilibrium that exhibit fluctuations-driven complex oscillations. We then delve into the analysis of entropy production rate (dissipation) in the various types of stochastic complex oscillations exhibited by these two nonlinear, nonequilibrium biological models. Our findings show that for these two models, complex oscillations have higher values of steady-state total entropy production rate $\dot{S}^{\mathrm{ss}}_{\mathrm{tot}}$ than simple periodic oscillations. Moreover, in the Hindmarsh-Rose neuronal model, we observe an order-to-disorder transition from periodic (organized) bursts of spikes to chaotic (unorganized) oscillations with distinct behaviors of $\dot{S}^{\mathrm{ss}}_{\mathrm{tot}}$. The results indicate that complex oscillations are produced at the cost of higher energy consumption and and that it requires a higher thermodynamic cost to maintain the periodic bursts than chaotic oscillations. The results obtained deepen our understanding of energy dissipation in nonlinear, nonequilibrium biological systems with stochastic complex oscillatory dynamics. 

The paper is structured as follows. In Sec.~\ref{sec:two}, we introduce nonequilibrium chemical reaction systems. Our main results and discussion are presented in Sec.~\ref{sec:three}. In Sec.~\ref{sec:threea}, we detail the stochastic modeling of the intracellular calcium oscillation and the scaled Hindmarsh-Rose oscillator models. In Sec.~\ref{sec:threeb}, we analyze dissipation in the two nonlinear, nonequilibrium models. Finally, our concluding remarks are provided in Sec.~\ref{sec:conclu}.
\section{Nonequilibrium Chemical reaction system}
\label{sec:two}
\begin{figure}[h]

\includegraphics[scale=0.25]{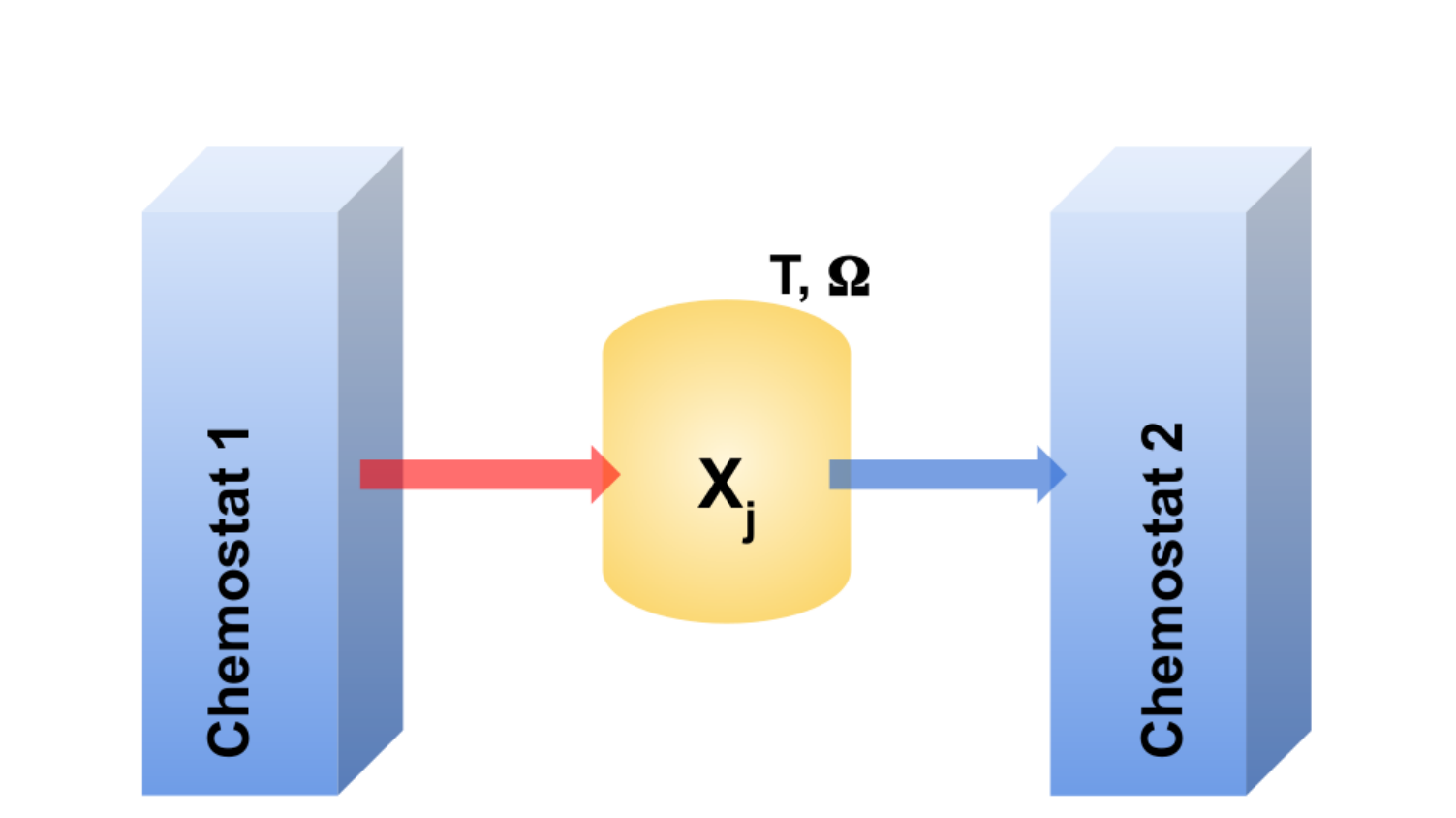}
\centering 
    \caption{\small{Open chemical reaction system far from thermodynamic equilibrium: the chemically reacting system (yellow cylinder) of system size $\Omega$ maintained at temperature $T$ is connected to two external chemostats or reservoirs (cuboids). In the reacting system, intermediate chemical species $X_j$ undergoes chemical reactions with the chemostat species $R_{j}^{I}$ supplied from the chemostat I (source), and the product(s) formed is removed to the chemostat II (sink), driving the system far from thermodynamic equilibrium. }}
    \label{fig:opensystem1}
\end{figure}
Suppose we have a well-stirred (space independence) chemically reacting system of a given system size $\Omega$ that is maintained at a fixed temperature $T$. By connecting the system to two external reservoirs or chemostats, one acting as a source and the other as a sink, the chemical reaction system transforms into an open system (See Fig.~\ref{fig:opensystem1}). Let $\textbf{X}=\{X_1,X_2,\dots \}$ and 
$\textbf{R}=\{R_{j}^{k}\}=\{R_{1}^{I},R_{2}^{I},\dots , R_{1}^{II},R_{2}^{II},\dots \}$ with $k\in I,II$ 
denote the sets of intermediate species $X_j$ and chemostat species $R_j^k$, respectively. The open chemically reacting system can then be represented by a CRN consisting of a set of $m$ irreversible reactions as:

\small{
\ce{\sigma^X_{1m}X_1 + \sigma^X_{2m}X_2 + ... + \sigma^{RI}_{1m}R_1^{I} + \sigma^{RI}_{2m}R_2^{I} + ... +\sigma^{RII}_{1m}R_1^{II} + \sigma^{RII}_{2m}R_2^{II} + ... ->[\eta_m] \rho^X_{1m}X_1 + \rho^X_{2m}X_2 + ... + \rho^{RI}_{1m}R_1^{I} + \rho^{RI}_{2m}R_2^{I} + ... +\rho^{RII}_{1m}R_1^{II} + \rho^{RII}_{2m}R_2^{II} + ... \,}
}

\normalsize {\noindent}In general, the stoichiometric matrix $\textbf{S}^i$ consists of elements $S_{jm}^i=\rho_{jm}^i-\sigma_{jm}^i$, where $i\in \{\textbf{X},\textbf{R}\}$, and $\sigma$ and $\rho$ are the stoichiometric coefficients for reactants and products, respectively. For an open reaction system, we consider the stoichiometric matrix $\textbf{S}=\textbf{S}^X$ accounting for changes solely in the intermediates $X_j$ in reaction $m$. The concentrations of $R_j^k$ chemostat species are assumed to be in significant excess compared to those of $X_j$ (i.e., $r_j^k\gg x_j$). We thus consider ideal chemostats maintaining constant $r_j^k$ with an overall chemical potential difference between chemostat I (source) and chemostat II (sink) denoted as $\Delta \mu\equiv \displaystyle \sum_j \mu_{R_{j}^{I}}- \sum_j \mu_{R_{j}^{II}}$. Species $R_j^k$ has the chemical potential $\mu_{R_j^k}=\mu_{R_j^k}^{0}+\ln r_j^k$, where $\mu_{R_j^k}^0$ denotes the chemical potential of the pure species~\cite{zhang2020dynamic}. The expression for $\Delta \mu$ will therefore be complicated due to the varying nature of the species involved. In the definition of the chemical potential $\mu_{R_j^k}$, we assume the product of gas constant and temperature, i.e., $RT=1$. For our analysis, we assume a non-zero $\Delta \mu$, which generates a flow of species from the source chemostat to the sink chemostat through a set of irreversible reactions in the open chemically reacting system of Fig.~\ref{fig:opensystem1}, thus performing chemical work and liberating free energy. The flow drives the system into nonequilibrium by breaking the detailed balance, resulting in nonequilibrium steady states (NESS) in the long time limit~\cite{qian2001mesoscopic}. An open chemical reaction system can give rise to dissipative structures of complex temporal and/or spatial order~\cite{prigogine1978time}, such as oscillations, waves, and patterns~\cite{lefever1971chemical}. In this study, we focus solely on temporal oscillations.

The reaction rate equations~\eqref{eq:rbnew22} describe the deterministic and continuous dynamics of concentrations in a chemical reaction system, assuming a large number of molecules or system size $\Omega$. However, in small systems such as biological cells, the populations of molecular species  (e.g., mRNA, proteins) change randomly in discrete integer amounts. This necessitates a framework with discrete space and continuous time, provided by stochastic modeling via the chemical Master equation (CME) (see Eq.~\eqref{eq:meqn} in Appendix~\ref{sec:appen0}). The CME models the time evolution of species' population as a Markov process, with $\mathcal{P}(\mathbf{X},t)$ denoting the probability of finding the chemical reaction system in state configuration $\mathbf{X}$ at time $t$. Alternatively, to account for the intrinsic fluctuation due to limited molecular number, one can use the Langevin equation to describe the stochastic dynamics of the chemical reaction system:
\begin{equation}
    \dot{\textbf{x}}=\textbf{F}(\textbf{x})+\textbf{G}\cdot \zeta,
\end{equation}
where the state variable $\textbf{x}=\textbf{X}/\Omega$ represents the species concentration, $\textbf{F}$ is the deterministic driving force, the matrix $\textbf{G}$ denotes the concentration-dependent part of the noise, and the vector $\zeta$ the time-dependent part of the noise~\cite{li2012potential}.

The Fokker-Planck (FP) equation from the continuous description of intrinsic fluctuations (see Appendix~\ref{sec:appen0}) is given by
\small
\begin{align}
&\frac{\partial P(\mathbf{x},t)}{\partial t}\nonumber \\ &= -\sum_{i}\frac{\partial}{\partial x_{i}}[f_{i}(\mathbf{x})P(\mathbf{x},t)] + \frac{1}{2\Omega}\sum_{i,j}\frac{\partial^{2}}{\partial x_{i}\partial x_{j}}[D_{ij}(\mathbf{x})P(\mathbf{x},t)] \nonumber \\
& =-\nabla \cdot \textbf{J}(\textbf{x},t). \label{eq:fpe}
\end{align}
\normalsize
From Eq.~\eqref{eq:fpe}, $\frac{\partial P(\mathbf{x},t)}{\partial t}+\nabla \cdot \textbf{J}(\textbf{x},t)=0$ represents a probability conservation equation where the local change in probability is equal to the net flux entering or outgoing. The deterministic drift is given by $\textbf{f}(\mathbf{x}) = \textbf{S} \cdot \nu(\mathbf{x} \cdot \Omega)$, with propensity function $\nu$ (see definition in Eq.~\eqref{eq:pro}). The diffusion matrix is given by
\begin{align}
\textbf{D}(\mathbf{x}) &= \textbf{S} \cdot \text{diag}(\nu) \cdot \textbf{S}^{\text{T}}, \label{eq:d}
\end{align}
 where $\textbf{D}$ is positive definite \cite{mendler2018analysis}. In Eq.~\eqref{eq:d}, $\textbf{S}^{\text{T}}$ denotes transposition of stoichiometric matrix $\textbf{S}$. 
 
 The probability current $\textbf{J}(\textbf{x},t)$, generated due to the irreversible species flow in the open reaction system of Fig.~\ref{fig:opensystem1}, is given by~\cite{kim2021thermodynamic}
\begin{equation}
    \textbf{J}(\mathbf{x}, t) = \textbf{H}(\mathbf{x})\mathbf{\alpha}(\mathbf{x}) P(\mathbf{x}, t) - \frac{1}{2\Omega} \textbf{D}(\mathbf{x}) \cdot \nabla P(\mathbf{x}, t). \label{eq:j}
    \end{equation} 
In the right-hand side of Eq.~\eqref{eq:j}, the first term denotes the convection current, and the second term the diffusion current~\cite{mendler2018analysis}. The matrix $\textbf{H}(\mathbf{x}) = \frac{1}{k_BT} \mathbf{D}(\mathbf{x})$ with Boltzmann constant $k_B$, and the term $\mathbf{\alpha}(\mathbf{x})$ is defined by
\begin{equation}
\mathbf{\alpha}(\mathbf{x}) = \textbf{f}(\mathbf{x}) - \frac{1}{2\Omega}\sum_{ik}\frac{\partial D_{ik}}{\partial x_{k}}\vec{\epsilon_{i}}, \label{eq:alpha}
\end{equation}
with the unity vector $\vec{\epsilon_{i}} = \frac{\vec{x_i}}{|x_i|}$ in the direction of $x_i$~\cite{mendler2018analysis}.

\noindent The probability current $\textbf{J}(\textbf{x},t)$ gives rise to energy dissipation, quantified by entropy production rate, in the chemically reacting system. For a nonequilibrium system described by the FP equation~\eqref{eq:fpe}, the system's entropy is defined as~\cite{seifert2019stochastic}
\begin{equation}
    S=-\int P(\textbf{x},t) \ln P(\textbf{x},t) \ d\textbf{x}.
\end{equation}
The rate of change in the system's entropy is~\cite{cao2015free}
\begin{align}
\dot{S}&=-\int \frac{\partial P(\textbf{x},t)}{\partial t} [\ln P(\textbf{x},t)+1] \ d\textbf{x} \nonumber \\
&=\int \frac{\textbf{J}^{\text{T}}(\mathbf{x}, t)\cdot {D}^{-1}(\mathbf{x})\cdot\textbf{J}(\mathbf{x}, t)}{P(\textbf{x},t)}\ d\mathbf{x}\nonumber \\
&-\int \textbf{J}^{\text{T}}(\mathbf{x}, t)\cdot{D}^{-1}(\mathbf{x})\cdot[\textbf{f}(\mathbf{x})-\nabla \cdot \textbf{D}(\textbf{x})] \ d\mathbf{x}\nonumber \\
&=\dot{S}_{\mathrm{tot}}-\dot{S}_{\mathrm{res}}.
\end{align}
The total entropy change of the system or the total entropy production rate (both system and reservoir) is thus given by  
\cite{cao2015free,kim2021thermodynamic}
 \begin{align}
     \dot{S}_{\mathrm{tot}}=\int \frac{\textbf{J}^{\text{T}}(\mathbf{x}, t)\cdot {D}^{-1}(\mathbf{x})\cdot \textbf{J}(\mathbf{x}, t)}{P(\textbf{x},t)} \ d\mathbf{x}\geq 0, \label{eq:epr}
 \end{align}
which implies the second law of thermodynamics. The entropy production rate can be considered aa a quantitative measure of the degree of detailed balance breaking or irreversibility of the nonequilibrium system~\cite{yan2023thermodynamic}. The term $\dot{S}_{\mathrm{res}}$ represents the entropy production rate from the reservoirs. In our analysis, we have used $k_BT = 1$ making entropy dimensionless.     

\section{Results and Discussion}
\label{sec:three}
\subsection{Stochastic Modeling: Formulation of Chemical Reaction Schemes of the Model Oscillators}
\label{sec:threea}
\subsubsection{Intracellular Calcium (Ca$^{2+}$) Oscillation Model}
\label{sec:model2}
Experimental studies have shown that calcium ions (Ca$^{2+}$) exhibit complex oscillations, such as bursting and quasi-periodic oscillations within biological cells. Houart \textit{et al.}~\cite{houart1999bursting} developed a nonlinear model based on the interplay between Ca$^{2+}$-induced Ca$^{2+}$ release (CICR)~\cite{puebla2005controlling} and Ca$^{2+}$-activated 1,4,5-trisphosphate (InsP$_3$) degradation mechanisms to explain these experimentally observed complex oscillations. If $x=\frac{X}{\Omega},~y=\frac{Y}{\Omega}$ and $z=\frac{Z}{\Omega}$ represent the concentrations of the free  Ca$^{2+}$ in the cytosol (cytosolic Ca$^{2+}$), Ca$^{2+}$ stored in the internal pool, and InsP$_3$, respectively, the intracellular calcium oscillation (ICO) model is described by the following set of coupled, nonlinear ODEs~\cite{houart1999bursting}:
\begin{align}
\dot{x}&=V_0+V_1\beta-V_2+V_3+k_fy-kx, \nonumber \\
\dot{y}&=V_2-V_3-k_fy, \label{eq:camodel}  \\
\dot{z}&=\beta V_4-V_5-\epsilon z, \nonumber 
\end{align}
where 
\begin{align}
 V_2&=V_{M2} \frac{x^2}{k_2^2+x^2},  \nonumber \\
 V_3&=V_{M3} \frac{x^m}{k_x^m+x^m} \frac{y^2}{k_y^2+y^2} \frac{z^4}{k_z^4+z^4},  \nonumber \\
 V_5&=V_{M5} \frac{z^p}{k_5^p+z^p} \frac{x^n}{k_d^n+x^n}. \nonumber
\end{align}
We explain the underlying mechanism and the parameters of the ICO model~\eqref{eq:camodel} in Appendix~\ref{app:A1} (see Ref.~\cite{chanu2024exploring} for details).  

We now proceed to formulate a stochastic chemical oscillator of the  ICO model~\eqref{eq:camodel} using the chemical Langevin equation (CLE), a stochastic differential equation (see Appendix~\eqref{app:Acle}). The concentration evolution equations $\dot{x}$, $\dot{y}$, and $\dot{z}$ of Eq.~\eqref{eq:camodel} are converted into a set of twelve reaction channels, detailed in the middle column of Table~\ref{table:two} in Appendix~\ref{app:A1}. Each reaction depicts the transitions of states of the species' populations $X,~Y$, and $Z$, representing random births and deaths of the species. The propensity function $\nu$ of each reaction step, calculated using Eq.~\eqref{eq:pro}, is detailed in the right column of Table~\ref{table:two}. In the propensity functions, $x= \frac{X}{\Omega},~y = \frac{Y}{\Omega}$, and $z= \frac{Z}{\Omega}$ represent the species concentrations with $\Omega$ as the system size. 



To determine the CLE~\cite{gillespie2000chemical} for the ICO model~\eqref{eq:camodel}, we first define the state vector $\textbf{x}=[x(t),y(t),z(t)]^{\mathrm{T}}$, where T implies transpose. Using the transitions and propensity functions detailed in Table~\ref{table:two}, we express the CLE for the ICO model~\eqref{eq:camodel} as:
\begin{equation}
d\textbf{x} = \boldsymbol{F}(\textbf{x})dt + \frac{1}{\sqrt{\Omega}} \boldsymbol{G}(\textbf{x})d\boldsymbol{W}. \label{eq:cle22} 
\end{equation}
Here, the function $\boldsymbol{F}(\textbf{x})$ is given by: 
\begin{align}
&\boldsymbol{F}(\textbf{x})=\begin{bmatrix}
& V_0+V_1\beta-V_2+V_3+k_fy-kx\\
    &V_2-V_3-k_fy\\
      &\beta V_4-V_5-\epsilon z\nonumber 
      \end{bmatrix}.
     \end{align}  
The stoichiometric matrix $\textbf{S}$ for the ICO model~\eqref{eq:camodel} is 
\begin{equation}
\textbf{S}=\left(
\begin{array}{cccccccccccc}
 1 & 1 & -1 & 1 & 1 & -1 & 0 & 0 & 0 & 0 & 0 & 0 \\
 0 & 0 & 0 & 0 & 0 & 0 & 1 & -1 & -1 & 0 & 0 & 0 \\
 0 & 0 & 0 & 0 & 0 & 0 & 0 & 0 & 0 & 1 & -1 & -1  \\
\end{array}
\right). \label{eq:stoi2}
\end{equation}
\normalsize 
Next, we calculate the matrix $\boldsymbol{G}(\textbf{x}) = \textbf{S}\cdot diag(\nu)$ using $\textbf{S}$ in Eq.~\eqref{eq:stoi2} and propensity functions $\nu$ listed in Table~\ref{table:two}. With the Wiener noise vector $\boldsymbol{dW}$, we finally obtain:
\begin{align}
     &\boldsymbol{G}(\textbf{x})d\boldsymbol{W}\nonumber \\
     &=\begin{bmatrix}
      \sqrt{V_0}\xi_1+\sqrt{V_1 \beta}\xi_2-\sqrt{V_2}\xi_3+\sqrt{V_3}\xi_4 +\sqrt{k_fy}\xi_5-\sqrt{kx}\xi_6\\
    \sqrt{V_2}\xi_7-\sqrt{V_3}\xi_8-\sqrt{k_fy}\xi_9\\
    \sqrt{V_4 \beta}\xi_{10}-\sqrt{V_5}\xi_{11}-\sqrt{\epsilon z}\xi_{12}
     \end{bmatrix}.  \label{eq:noise}
     \end{align}    
\normalsize
In Eq.~ \eqref{eq:noise}, $\xi_j \ ; \ j=1,2,\dots,12$ are statistically independent Gaussian white noises with properties $\langle \xi_j (t)\rangle=0$ and $\langle \xi_i(t) \xi_{j}(t')\rangle=\delta_{ij} \delta(t-t').$ 

To relate with the deterministic ODEs~\eqref{eq:camodel}, we note that the function $\boldsymbol{F}(\textbf{x})$ in the first term of the right-hand side of Eq.~\eqref{eq:cle22} corresponds to these deterministic ODEs. The second term accounts for the intrinsic fluctuations arising from random births and deaths of species in the chemical reaction system. The intrinsic fluctuation scales inversely with the square root of the system size as $\sim \frac{1}{\sqrt{\Omega}}$. As $\Omega$ increases, intrinsic fluctuation decreases. In the large limit $\Omega \rightarrow \infty$, the system dynamics shows a stochastic to deterministic transition, known as thermodynamic limit. 

Solving the CLE~\eqref{eq:cle22} of the ICO model~\eqref{eq:camodel} using standard modules such as \texttt{solve\_ivp} (from \texttt{SciPy} package in Python) for different parameter values of $\epsilon$ and $\beta$~\cite{houart1999bursting,chanu2024exploring} reveals diverse dynamics in the concentration $x(t)$ of cytosolic Ca$^{2+}$. The panel (a) in Figure~\ref{samplee} illustrates $x(t)$ exhibiting simple periodic oscillations (Top), quasi-periodic oscillations (Middle), and bursting (Bottom). Simple periodic oscillation entails a single frequency or time period, whereas quasi-periodic oscillations involve multiple frequencies. The limit cycle for quasi-periodic oscillations undergoes a secondary Hopf bifurcation, covering a torus~\cite{houart1999bursting}. On the other hand, bursting is characterized by a single sharp spike followed by smaller spikes of reduced amplitudes around a plateau. For bursting, the limit cycle emerges through a Hopf bifurcation~\cite{houart1999bursting}. For in-depth characterizations of the deterministic counterparts of these stochastic complex oscillations with methods such as first return maps, power spectra, and Lyapunov exponents, we refer the readers to Ref.~\cite{houart1999bursting}.

Again, we see that all the reactions underlying the ICO model~\eqref{eq:camodel} in Table~\ref{table:two} of Appendix~\ref{app:A1} are irreversible and far from thermodynamic equilibrium. We analyze the rate of entropy production (dissipation) of this nonlinear, nonequilibrium system in Sec.~\ref{sec:threeb}.

\subsubsection{Scaled Hindmarsh-Rose (sHR) Oscillator Model }
\label{sec:model1}

\begin{figure}
\includegraphics[scale=0.4]{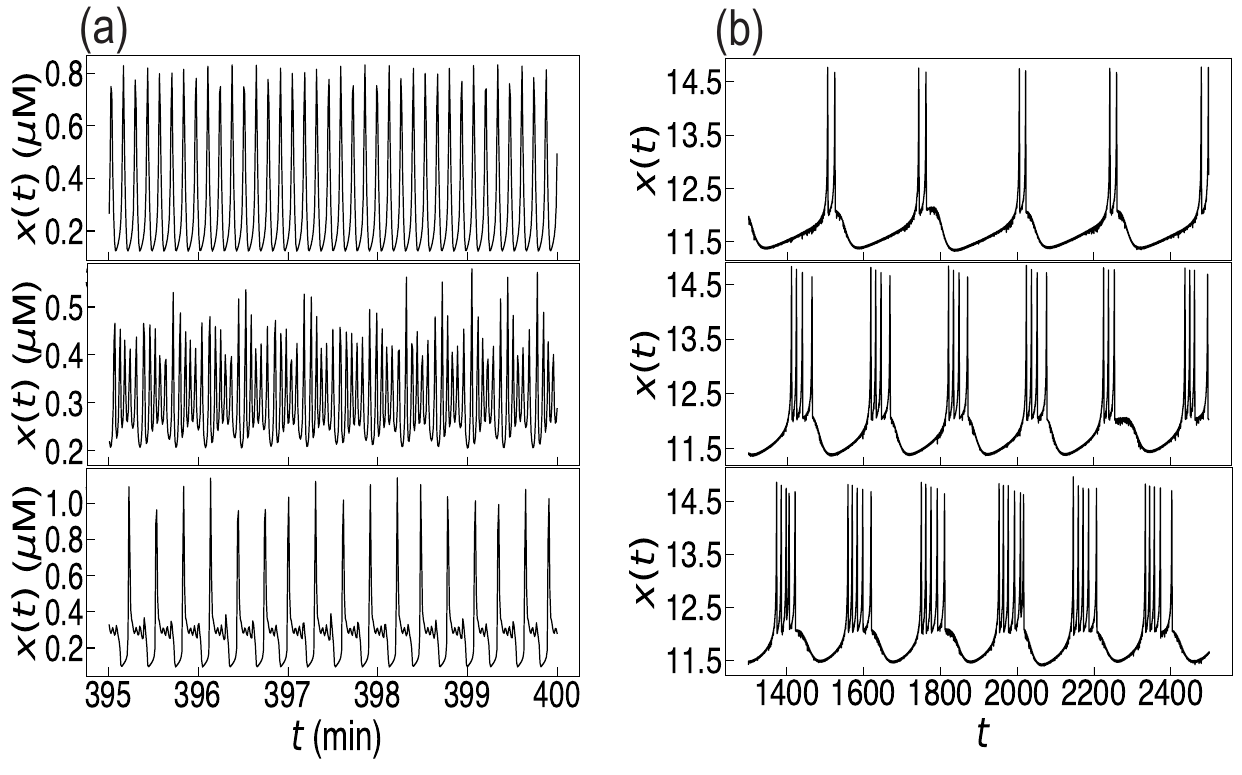}
\centering 
    \caption{\small{Plots of concentrations $x(t)$ versus time $t$: (a) Simple periodic oscillations (Top), quasi-periodic oscillations (Middle), and bursting (Bottom) are obtained by solving the CLE~\eqref{eq:cle22} of the intracellular calcium (Ca$^{2+}$) oscillation (ICO) model~\eqref{eq:camodel}. The cytosolic Ca$^{2+}$ concentration $x(t)$ is measured in $\mu$M and $t$ in minutes (min). (b) Bursts of spikes in $x(t)$ are obtained by solving the CLE~\eqref{eq:cle2} for the scaled Hindmarsh-Rose oscillator model~\eqref{eq:shrm}. Periodicity $p$ equals 2 (Top), 4 (Middle), and 5 (Bottom). $x(t)$ and $t$ are in arbitrary units.}}
    \label{samplee}
\end{figure}

\begin{figure}
\includegraphics[scale=0.35]{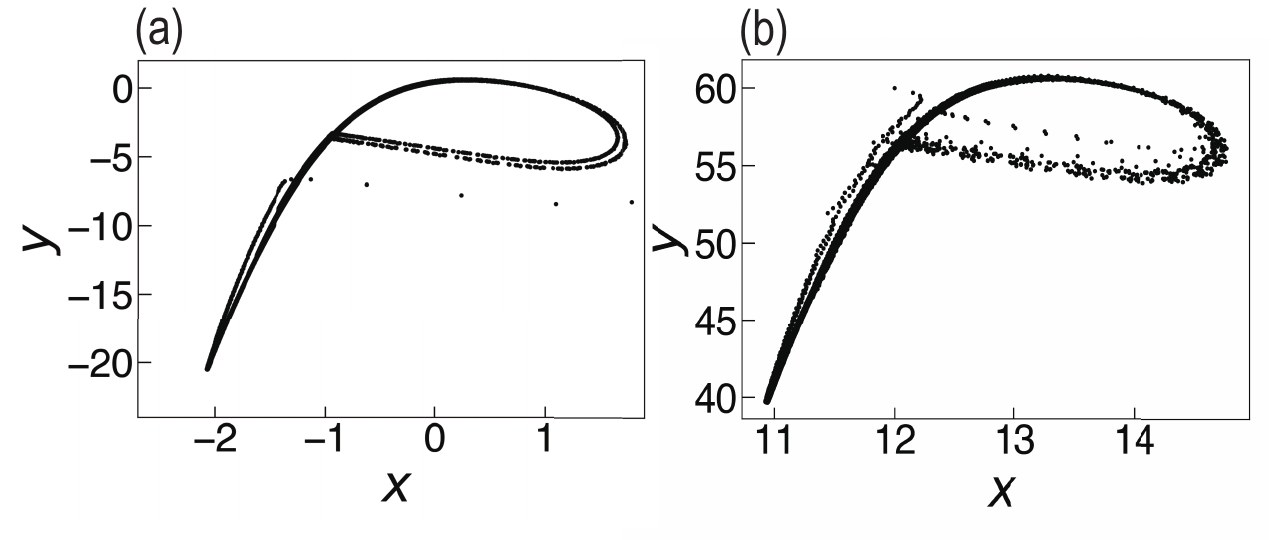}
    \caption{\small{$x-y$ phase plots: (a) original Hindmarsh-Rose model of Eq.~\eqref{eq:hrm}, and (b) modified Hindmarsh-Rose model of Eq.~\eqref{eq:shrm} after scaling. The parameters used are $I=1.44$, and $r=0.003$ with initial conditions $x_0=12,~y_0=60$, and $z_0=3$. For (b), scaling parameters $\alpha_x=13$ and $\alpha_y=60$.}}
    \label{scale}
\end{figure}

The nonlinear Hindmarsh-Rose (HR) model~\cite{hindmarsh1982model,hindmarsh1984model} is capable of explaining diverse bursting phenomena exhibited by neuronal cells. Described in Appendix~\ref{app:A1}, the HR neuronal model is a three-variable model represented by the state vector $\mathbf{x}=[x(t),y(t),z(t)]^{\text{T}}$ (See Eq.~\eqref{eq:hrm}). The solutions of the HR model~\eqref{eq:hrm} span the entire $\mathbb{R}^3$~\cite{hindmarsh1984model}. We now propose reformulating the HR model~\eqref{eq:hrm} as a stochastic chemical oscillator. Since we are now considering species concentrations, we require $\mathbf{x} > 0$. To ensure positivity of $\mathbf{x}$, we follow the scaling method provided by Poland~\cite{poland1993cooperative}, where we take $x \rightarrow x+\alpha_x$ and $y \rightarrow y+\alpha_y$ in Eq.~\eqref{eq:hrm}. Thus we obtain the \textit{scaled} Hindmarsh-Rose (sHR) oscillator model as:
\begin{align}
   \dot{x}&=(y - \alpha_{y}) + 3(x - \alpha_{x})^2 - (x - \alpha_{x})^3 - z + I,\nonumber \\
      \dot{y}&=1 - 5(x - \alpha_{x})^2 - (y - \alpha_{y}), \label{eq:shrm}\\
       \dot{z}&=4r(x - \alpha_{x}) + \frac{32r}{5} - rz. \nonumber 
     \end{align}  
We choose the scaling parameters $\alpha_{x} = 13 $ and $\alpha_{y} = 60$. Since $z(t)>0$ $\forall \ t$, no scaling in $z$ is needed. The scaling shifts the center of the limit cycle obtained by solving Eq.~\eqref{eq:shrm}, moving the dynamics into the first quadrant region where $\mathbf{x}>0$, as depicted in the $x$--$y$ phase plots of Fig.~\ref{scale}. Scaling does not affect the overall dynamical behavior. 

We now proceed to formulate a stochastic chemical oscillator of the sHR model~\eqref{eq:shrm} using the CLE. We first convert the evolution equations $\dot{x}$, $\dot{y}$, and $\dot{z}$ of Eq.~\eqref{eq:shrm} into a set of sixteen reaction channels, detailed in the middle column of Table~\ref{table:one} in Appendix~\ref{app:A1}. The propensity function $\nu$ of each reaction step is calculated using Eq.~\eqref{eq:pro}, listed in the right column of Table~\ref{table:one}.

We now elucidate the chemical reaction mechanisms of the reaction channels listed in Table~\ref{table:one}. The coupled slow-fast reaction schemes result from specific coupling situations in the ODEs~\eqref{eq:shrm}. To illustrate, let us consider the term $-z$ in $\frac{dx}{dt}$. This term can be realized through two reaction steps by coupling a slow reaction with a very fast one: \ce{D_2 +Z -> D_2^* +Z (slow)}, \ce{ D_2^* +X -> D_2 + R (fast)}. In the first slow reaction, a catalytic species \ce{D_2} interacts with \ce{Z} to produce the activated \ce{ D_2^*} species. Subsequently, \ce{ D_2^*} combines with \ce{X}, and the resulting products are removed by the sink in a very fast reaction. Essentially, species $Z$ catalyzes the annihilation of $X$, with the reaction rate primarily determined by the slow reaction step. Another example that involves the slow-fast couples is \ce{D_5 +X -> D_5^* +X (slow)} and \ce{D_5^* +B -> Y + D_5 (fast)}. Here, $X$ catalyzes the production of $Y$. Additionally, reactions like \ce{X -> R}, or \ce{Y -> R} represent irreversible sink reactions. In the formulated reaction schemes, higher-order reaction mechanisms are present, for instance, the second reaction \ce{3X -> R} implies a third-order sink reaction (trimolecular reaction kinetics), while the third reaction \ce{2X -> 3X} indicates a second-order autocatalytic reaction  (bimolecular chemical kinetics).

For the sHR oscillator model~\eqref{eq:shrm}, we determine the stoichiometric matrix $\textbf{S}$ as: 
\tiny{
\begin{equation}
\textbf{S}=\left(
\begin{array}{cccccccccccccccc}
 1 & -3 & 1 & -1 & -1 & 1 & -1 & 0 & 0 & 0 & 0 & 0 & 0 & 0 & 0 & 0 \\
 0 & 0 & 0 & 0 & 0 & 0 & 0 & 1 & -1 & 1 & -1 & -1 & 0 & 0 & 0 & 0 \\
 0 & 0 & 0 & 0 & 0 & 0 & 0 & 0 & 0 & 0 & 0 & 0 & 1 & -1 & -1 & 1 \\
\end{array}
\right) \label{eq:stoi1}
\end{equation}
} 
\normalsize 
For the sHR oscillator model~\eqref{eq:shrm}, we define the state vector $\textbf{x}=[x(t),y(t),z(t)]^{\mathrm{T}}$, where T implies transpose. Using the transitions and propensity functions detailed in Table~\ref{table:one}, we express the CLE for the sHR oscillator model~\eqref{eq:shrm} as:
\begin{equation}
d\textbf{x} = \boldsymbol{F}(\textbf{x})dt + \frac{1}{\sqrt{\Omega}} \boldsymbol{G}(\textbf{x})d\boldsymbol{W}, \label{eq:cle2} 
\end{equation}
where $\boldsymbol{F}(\textbf{x})$ and $\boldsymbol{G}(\textbf{x})d\boldsymbol{W}$ are given by: 
\begin{align}
&\boldsymbol{F}(\textbf{x})=\begin{bmatrix}
&(y - \alpha_{y}) + 3(x - \alpha_{x})^2 - (x - \alpha_{x})^3 - z + I\\
    &1 - 5(x - \alpha_{x})^2 - (y - \alpha_{y})\\
      &4r(x - \alpha_{x}) + \frac{32r}{5} - rz \nonumber 
      \end{bmatrix},
     \end{align}  

\begin{widetext}
    \small
\begin{align}
     &\boldsymbol{G}(\textbf{x})d\boldsymbol{W}\nonumber \\
     &=\begin{bmatrix}
      \sqrt{y} \xi_{1} - \sqrt{3x^3} \xi_{2} + \sqrt{3(\alpha_{x}+1)x^2}  \xi_{3}  -  \sqrt{3\alpha_{x}x(2+\alpha_{x})}  \xi_{4} - \sqrt{z}  \xi_{5} + \sqrt{3\alpha_{x}^2 + \alpha_{x}^3 + I}  \xi_{6} - \sqrt{\alpha_{y}}  \xi_{7}\\
      \sqrt{1+\alpha_y} \xi_{8} - \sqrt{5x^2} \xi_{9} + \sqrt{10\alpha_{x}x} \xi_{10}  -  \sqrt{y}  \xi_{11} - \sqrt{5\alpha_{x}^2 }  \xi_{12}\\
     \sqrt{4rx}  \xi_{13} - \sqrt{rz}  \xi_{14} - \sqrt{4r\alpha_{x}}\xi_{15} + \sqrt{\frac{32r}{5}}  \xi_{16} \label{eq:g1}
     \end{bmatrix}.
     \end{align}    
     \end{widetext}
\normalsize

 
Solving the CLE~\eqref{eq:cle2} for the sHR oscillator model~\eqref{eq:shrm}  across various parameter values of $I$ and $r$, we obtain complex dynamics in $x(t)$, such as bursts of spikes with varying periodicities and chaos. The panel (b) of Figure~\ref{samplee} depicts plots of $x(t)$ versus time $t$ showing complex patterns of bursts of spikes: 2-period (Top), 4-period (Middle), and 5-period (Bottom). $x(t)$ and $t$ are in arbitrary units. Bifurcation analyses of such complex oscillations in the deterministic three-dimensional Hindmarsh-Rose model~\eqref{eq:hrm} are found in previous studies, see e.g. Refs.~\cite{gonzalez2007complex,storace2008hindmarsh,dtchetgnia2013deterministic}, providing insights into the system's nonlinear dynamics.

The reactions governing the sHR oscillator model~\eqref{eq:shrm}, detailed in Table \ref{table:one}, are all irreversible and far from thermodynamic equilibrium. In the next Sec.~\ref{sec:threeb}, we delve into the analysis of dissipation, quantified by entropy production rate, in this nonlinear, nonequilibrium system of sHR oscillator~\eqref{eq:shrm}.

\subsection{Dissipation in the stochastic chemical oscillators with complex oscillations}
\label{sec:threeb}
\begin{figure*}
\includegraphics[scale=0.5]{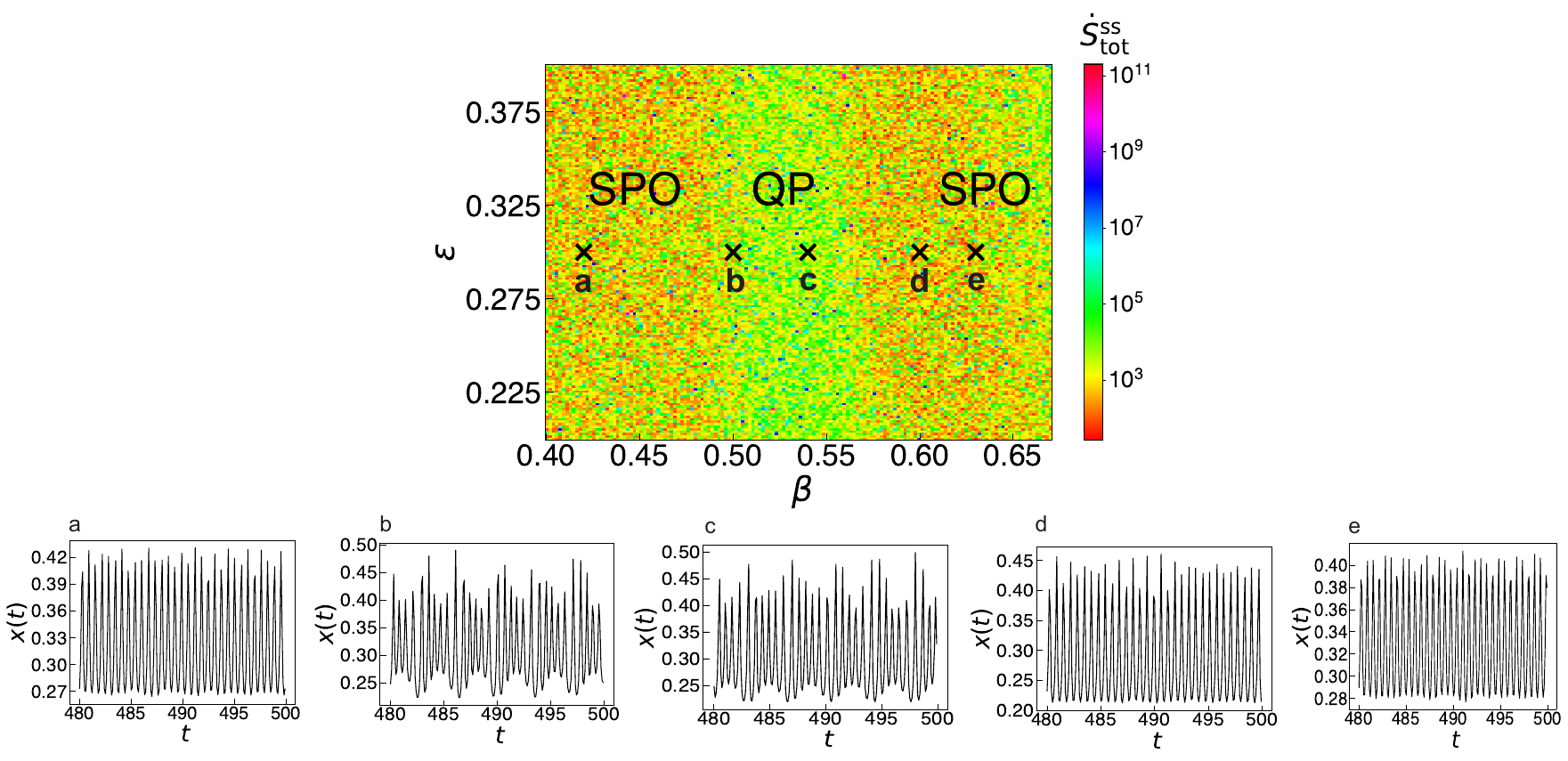}
\centering 
    \caption{\small{The heat map of total entropy production rate at steady-state ($\dot{S}^{\mathrm{ss}}_{\mathrm{tot}}$) of the intracellular calcium (Ca$^{2+}$) oscillation model~\eqref{eq:camodel} in the two-dimensional parameter space of $(\beta,\epsilon)$ where a region of quasi-periodic (QP) oscillations is embedded in a region of simple periodic oscillations (SPO). Markers $a$--$e$ illustrate the stochastic time series $x(t)$ of the cytosolic Ca$^{2+}$ concentration. $a,~d,~e$ present SPO, whereas $b,c$ display QP oscillation.}}
    \label{fig:ICOoscillations1}
\end{figure*}
\begin{figure*}
\includegraphics[scale=0.5]{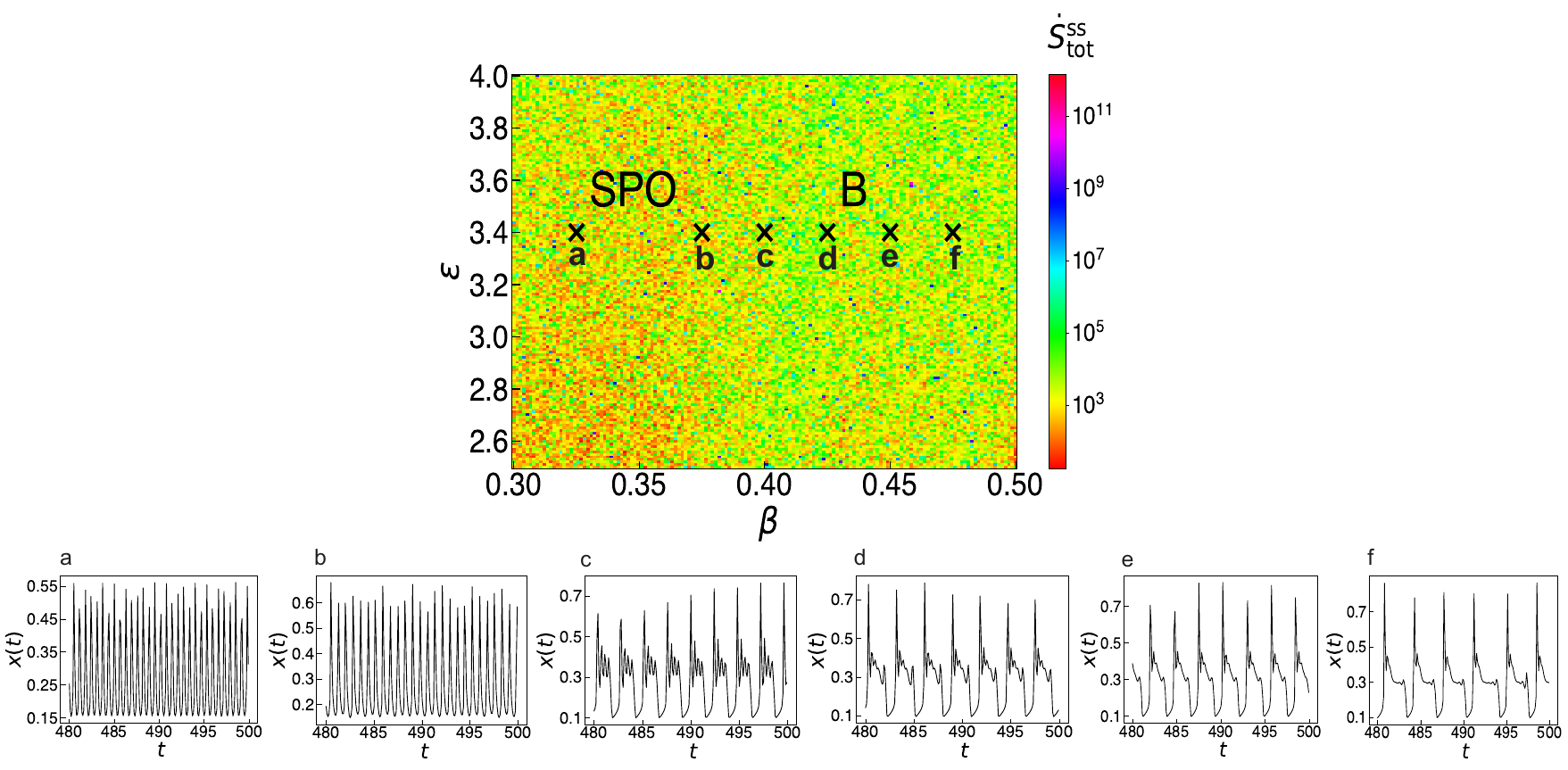}
\centering 
    \caption{\small{The heat map of $\dot{S}^{\mathrm{ss}}_{\mathrm{tot}}$ of the intracellular Ca$^{2+}$ oscillation model~\eqref{eq:camodel} in the $(\beta,\epsilon)$ parameter space where simple periodic oscillation (SPO) and bursting (B) are observed. The cytosolic Ca$^{2+}$ concentration $x(t)$ show SPO (markers $a,~b$) and bursting ($c-f$).}}
    \label{fig:ICOoscillations2}
\end{figure*}
\begin{figure*}
\includegraphics[scale=0.4]{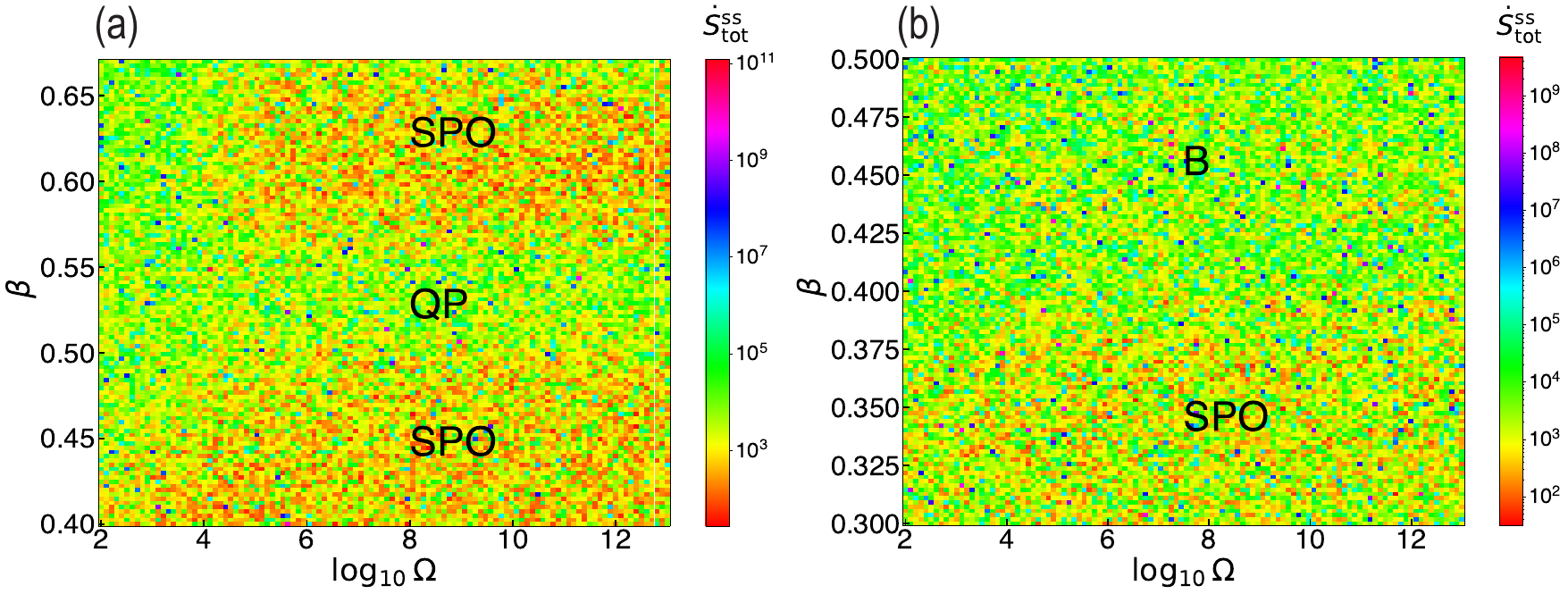}
    \caption{\small{Effect of system size ($\Omega$) on $\dot{S}^{\mathrm{ss}}_{\mathrm{tot}}$ of the intracellular Ca$^{2+}$ oscillation model~\eqref{eq:camodel} illustrated in the parameter space of ($\Omega,\beta$) corresponding to (a) $\epsilon=0.3$ in Fig.~\eqref{fig:ICOoscillations1}, and (b) $\epsilon=3.4$ in Fig.~\eqref{fig:ICOoscillations2}. The x-axes are in the log scale.}}
    \label{ssf}
\end{figure*}
Using the irreversible chemical reaction channels outlined in Tables~\ref{table:two}$\&$\ref{table:one} of Appendix~\ref{app:A1}, we now illustrate open chemical reaction systems for both the ICO model~\eqref{eq:camodel} and the sHR oscillator model~\eqref{eq:shrm}. For the ICO model~\eqref{eq:camodel},  the intermediate species $\textbf{X}=\{X,Y,Z\}$ and the chemostat species $\textbf{R}=\{A,B,\dots,K,M,M^{*}\}$. In the case of the sHR oscillator model~\eqref{eq:shrm}, $\textbf{X}=\{X,Y,Z\}$, and $\textbf{R}=\{A,B,C,D_1,\dots,D_8,D_1^{*},\dots,D_8^{*},R\}$. Using these species, for the two nonlinear models, we construct open reaction systems as shown in Fig.~\ref{fig:opensystem1} that generate a flow of species from $R^I$ chemostat (acting as a source) to another chemostat $R^{II}$ (serving as a sink) through the set of irreversible reactions (detailed in Tables~\ref{table:two}$\&$\ref{table:one} of Appendix~\ref{app:A1}) involving intermediate species $\textbf{X}$, driving the systems into regimes of non-equilibrium. 

 We consider long-time steady-state solutions of the corresponding FP equations~\eqref{eq:fpe} of the ICO model~\eqref{eq:camodel} and the sHR oscillator model~\eqref{eq:shrm}. At the steady-state, $\frac{\partial P(\mathbf{x},t)}{\partial t}=0$, resulting in the steady-state probability distribution $P^{\mathrm{ss}}(\textbf{x})$. From the steady-state condition of $\nabla \cdot \textbf{J}(\textbf{x},t)=0$, two possible solutions of the steady-state probability current $\textbf{J}^{\mathrm{ss}}(\mathbf{x})$ arise: (1) $\textbf{J}^{\mathrm{ss}}(\mathbf{x})=0$ (satisfying detailed balance condition implying the system being in equilibrium), and (2) $\textbf{J}^{\mathrm{ss}}(\mathbf{x})\neq 0$. In this second case, a non-zero flux indicates a net probability flux breaking the detailed balance and leading the system to nonequilibrium. For the open chemical reaction systems of the two models considered, in the long time-limit, we eventually reach the nonequilibrium steady states (NESS) with non-zero $\textbf{J}^{\mathrm{ss}}(\mathbf{x})$, giving rise to a local thermodynamic force $\mathcal{F}(\mathbf{x})$, given by $\mathcal{F}(\mathbf{x})=\frac{\textbf{J}^{\mathrm{ss T}}(\mathbf{x}).D^{-1}(\mathbf{x})}{P^{\mathrm{ss}}(\mathbf{x})}$~\cite{qian2001mesoscopic}. Here, T stands for transpose. The product of $\mathcal{F}(\mathbf{x})$ and $\textbf{J}^{\mathrm{ss}}(\mathbf{x})$ yields the local entropy production rate, given by $\dot{\sigma}^{\mathrm{ss}}(\mathbf{x})=\mathcal{F}(\mathbf{x})\textbf{J}^{\mathrm{ss}}(\mathbf{x})$, for the state $\mathbf{x}$. The total entropy production rate at steady-state, denoted by $\dot{S}^{\mathrm{ss}}_{\mathrm{tot}}$, is obtained by integrating $\dot{\sigma}^{\mathrm{ss}}(\mathbf{x})$ over all configurations $\mathbf{x}$. For simple systems like the bead-spring model~\cite{li2019quantifying}, one can derive an analytical expression for $\dot{S}^{\mathrm{ss}}_{\mathrm{tot}}$. However, for complex nonlinear systems like the ICO model~\eqref{eq:camodel} and the sHR oscillator model~\eqref{eq:shrm}, analytical solutions of $\dot{S}^{\mathrm{ss}}_{\mathrm{tot}}$ are rather difficult to find. Instead, we numerically determine $\dot{S}^{\mathrm{ss}}_{\mathrm{tot}}$ by estimating it directly from the time series $\textbf{x}(t)$ from steady-state distributions. 

To obtain the steady-state probability currents $\textbf{J}^{\text{ss}}(\textbf{x})$, we first determine the deterministic drift $\textbf{f}$, diffusion matrix $\textbf{D}$, and the term $\alpha$ for the ICO model~\eqref{eq:camodel} and the sHR oscillator model~\eqref{eq:shrm} (see the expressions in Appendix~\ref{app:A1}). Then, we determine the steady-state probability distribution $\hat{P}^{\mathrm{ss}}(\textbf{x})$ from the trajectories $\textbf{x}(t)$ obtained by numerically integrating the corresponding CLEs~\eqref{eq:cle22} and \eqref{eq:cle2}. We employ the method of estimating the multivariate probability distribution from single long-time trajectories~\cite{li2019quantifying}. Using Eq.~\eqref{eq:j}, we estimate the steady-state probability current $\hat{J}^{\mathrm{ss}}(\textbf{x})$ with the obtained $\hat{P}^{\mathrm{ss}}(\textbf{x})$. Finally, we estimate the total entropy production rate at steady-state ($\dot{S}^{\mathrm{ss}}_{\mathrm{tot}}$) using Eq.~\eqref{eq:epr}. The algorithmic pseudo-code to determine $\dot{S}^{\mathrm{ss}}_{\mathrm{tot}}$ is detailed in the \textbf{Methods} section of Appendix~\ref{app:method}. 

Firstly, we analyze the nonequilibrium system of the intracellular Ca$^{2+}$ oscillation (ICO) model~\eqref{eq:camodel}. We explore $\dot{S}^{\mathrm{ss}}_{\mathrm{tot}}$ with respect to the parameters $\beta$ and $\epsilon$, representing the degree of cellular stimulation by an agonist (e.g., ATP or hormone) and the degradation of InsP$_3$ by 5-phosphatase, respectively. We now investigate $\dot{S}^{\mathrm{ss}}_{\mathrm{tot}}$ in two distinct parameter space of $(\beta,\epsilon)$: (1) $\epsilon\in [0.2,0.4]$ and $\beta\in [0.40,0.65]$, where a quasi-periodic (QP) oscillatory domain is embedded within a region of simple periodic oscillations (SPO) (see Fig.~5b of Ref.~ \cite{houart1999bursting}), and (2) $\epsilon \in [2.5,4.0]$ and $\beta \in [0.3,0.5]$, where SPO and bursting (B) are observed (see Fig.~5a of Ref.~ \cite{houart1999bursting}). In these parameter domains, the open system of the ICO model~\eqref{eq:camodel} displays complex oscillations like quasi-periodic oscillations and bursting. These complex oscillations are linked to the regulatory mechanisms underlying  Ca$^{2+}$ oscillations that have physiological significance \cite{borghans1997complex,houart1999bursting}.

Figs.~\ref{fig:ICOoscillations1}$\&$\ref{fig:ICOoscillations2} illustrate heat maps of $\dot{S}^{\mathrm{ss}}_{\mathrm{tot}}$ corresponding to the two parameter domains mentioned above. Each $(\beta,\epsilon)$ grid contains $150\times 150$ data points of $\dot{S}^{\mathrm{ss}}_{\mathrm{tot}}$. In Fig.~\ref{fig:ICOoscillations1}, $\dot{S}^{\mathrm{ss}}_{\mathrm{tot}}$ of QP oscillations exhibits higher values than those of the embedding regions of SPO. This is evident from the markers $a$--$e$ depicting the cytosolic Ca$^{2+}$ concentration $x(t)$ for varying $\beta$ at a fixed $\epsilon=0.30$. Markers $a,~d$ and $e$ display SPO, while $b,~c$ exhibit QP oscillations. Similarly, in Fig.~\ref{fig:ICOoscillations2}, $\dot{S}^{\mathrm{ss}}_{\mathrm{tot}}$ of bursting (B) shows higher values than those of SPO. At a fixed $\epsilon=3.4$, markers $a,~b$ display SPO, while $c$--$f$ depict bursting dynamics of $x(t)$. The results in Figs.~\ref{fig:ICOoscillations1}$\&$\ref{fig:ICOoscillations2} demonstrate a clear dependence of $\dot{S}^{\mathrm{ss}}_{\mathrm{tot}}$ on different oscillatory dynamics, with higher entropy production rates observed in complex oscillations such as quasi-periodic oscillations and bursting compared to simple periodic oscillations.  

\begin{figure*}
\includegraphics[scale=0.6]{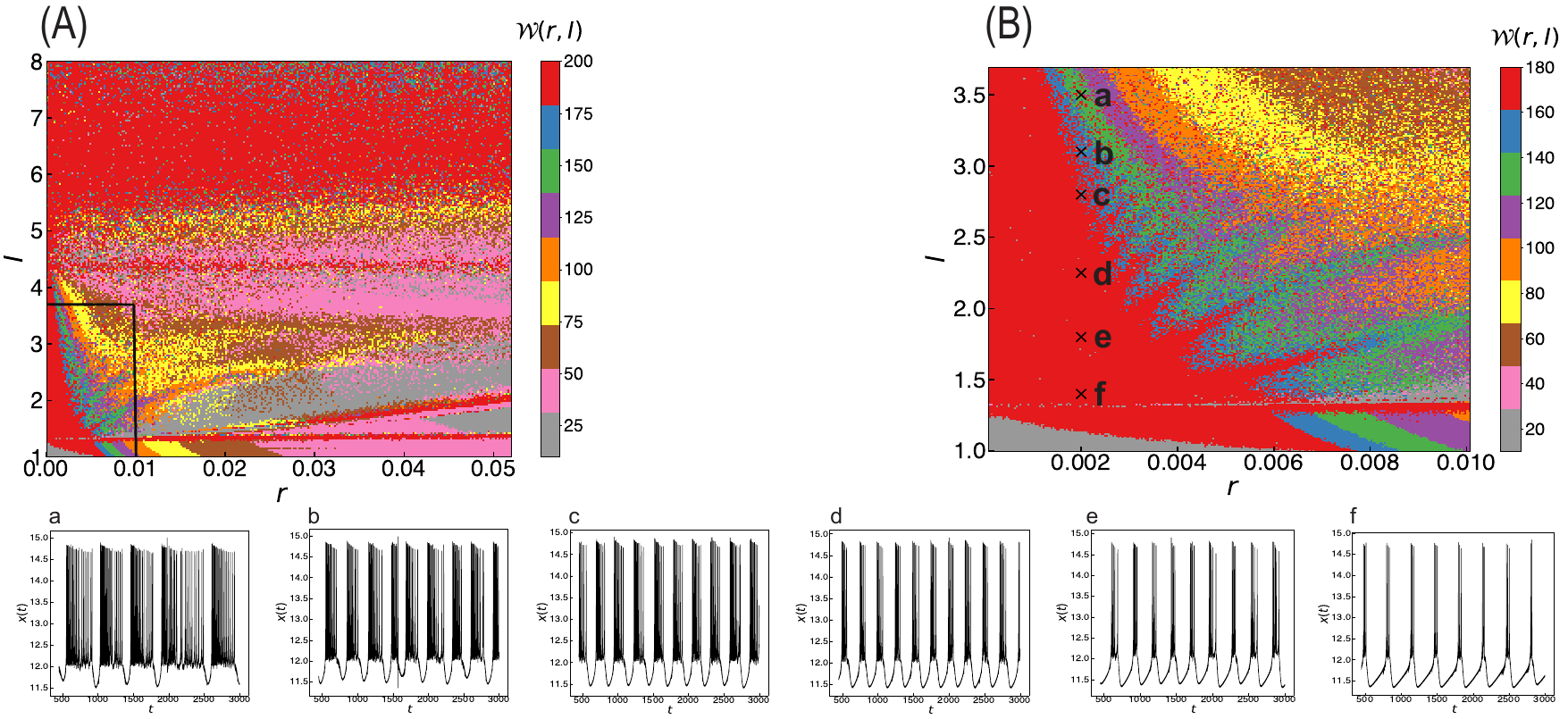} 
\caption{\small{Characterization of stochastic complex oscillations of the scaled Hindmarsh-Rose oscillator model~\eqref{eq:shrm} using a function $\mathcal{W}(r,I)$ (see Main text for definition), which measures inter-spike intervals (ISI). (A) The heat map of $\mathcal{W}(r,I)$ indicates regions of varying ISI, suggesting variations in the periodicity of the oscillations of $x(t)$ across $(r,I)$. (B) The heat map of $\mathcal{W}(r,I)$ for the smaller parameter region (marked by a rectangular box in Fig.~\ref{hrm1}(A)) where stochastic complex oscillations of bursts of spikes are observed, as illustrated by the time series $x(t)$ in markers $a$--$f$.}}
    \label{hrm1}
\end{figure*}
We investigate the effect of system size $\Omega$ on $\dot{S}^{\mathrm{ss}}_{\mathrm{tot}}$ of the ICO model~\eqref{eq:camodel} in the two-dimensional parameter space of $(\Omega,\beta)$, as shown in Figs.~\ref{ssf}(a)$\&$(b). We have taken $\epsilon=0.3$ and $3.4$ in Figs.~\ref{ssf}(a)$\&$(b), respectively. $\dot{S}^{\mathrm{ss}}_{\mathrm{tot}}$ for SPO, QP oscillation, and B reach saturation values for $\Omega \gtrsim 10^5$, indicating transition towards the thermodynamic limit.

We now proceed to analyze the nonequilibrium system of the sHR oscillator model~\eqref{eq:shrm} in the two-dimensional parameter space $(r,I)$. The parameter $I$ represents the external current entering the neuron and influencing its excitability patterns, while $r$ denotes the efficiency of slow ionic channels (chemically activating-deactivating), impacting the timescale and dynamics of neuronal firing~\cite{gonzalez2007complex}. These parameters allow a wide range of excitability patterns and firing regimes of $x(t)$ in the sHR oscillator model~\eqref{eq:shrm}.

To characterize the complex oscillations of the sHR oscillator model~\eqref{eq:shrm}, we consider a function $\mathcal{W}$~\cite{gonzalez2007complex}. Denoting $\mathcal{T}_{\mathrm{max}}$ and $\mathcal{T}_{\mathrm{min}}$ as the maximum and minimum of the inter-spike interval (time between consecutive peaks) $\mathcal{T}$ of the time series $x(t)$, the function $\mathcal{W}$ is defined as $\mathcal{W} = \mathcal{T}_{\mathrm{max}}- \mathcal{T}_{\mathrm{min}}$. $\mathcal{W}$ provides a measure of the variation of bursts of spikes in the sHR oscillator model~\eqref{eq:shrm}. $\mathcal{W}$ serves as a meaningful quantity that encapsulates how neurons encode messages in the patterns of inter-spike intervals~\cite{gonzalez2007complex}. Over a range of parameter space $(r,I)$, we analyze the complex patterns of bursts in $x(t)$ by computing $\mathcal{T}_{\mathrm{min}}(r,I), \mathcal{T}_{\mathrm{max}} (r,I)$ and hence $\mathcal{W}(r,I)$. 

\begin{figure*}
\includegraphics[scale=0.2]{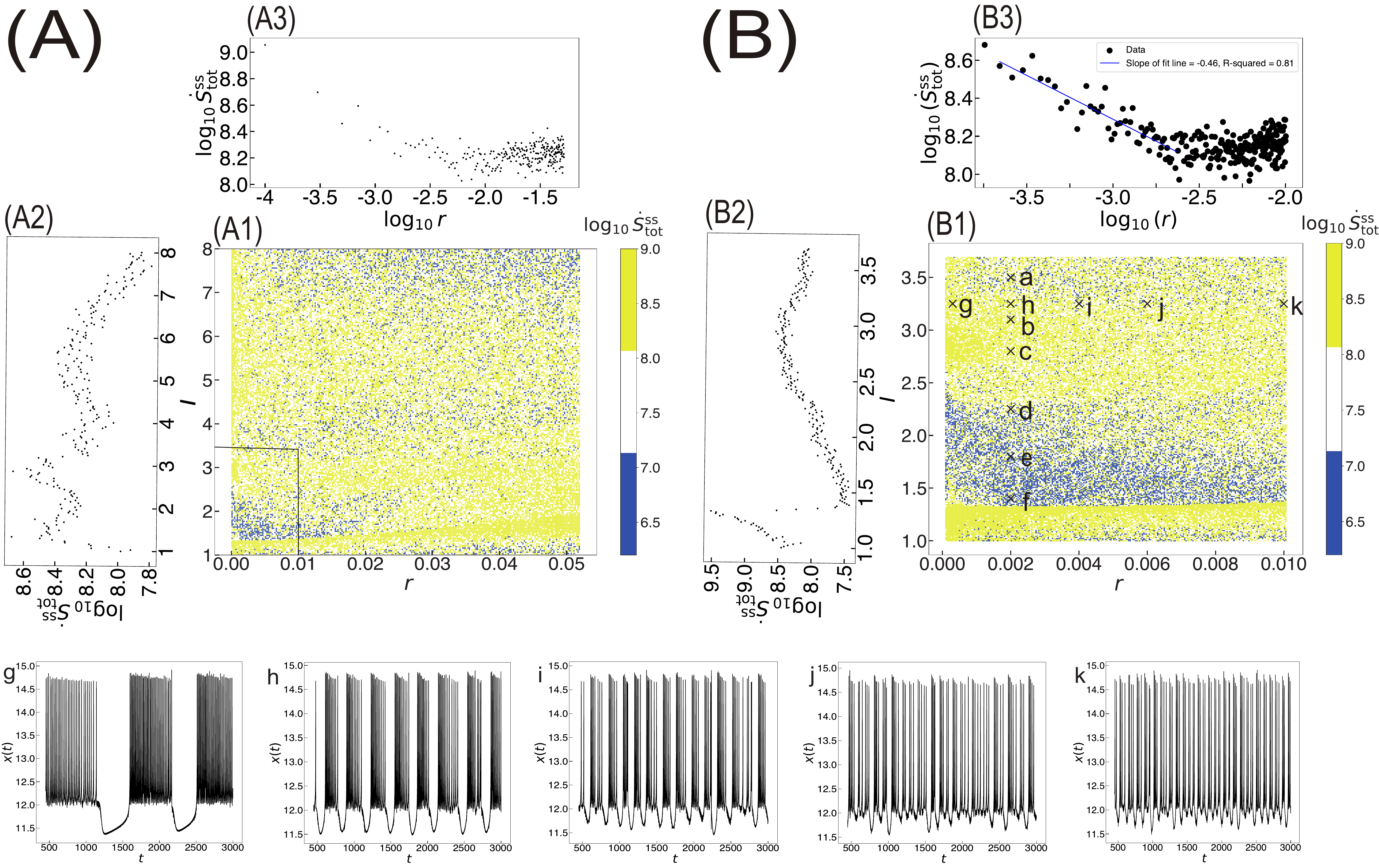}
    \caption{\small{The total entropy production rate at steady-state, $\dot{S}^{\mathrm{ss}}_{\mathrm{tot}}$ of the nonequilibrium scaled Hindmarsh-Rose oscillator model~\eqref{eq:shrm}. (A1) The heat map of $\dot{S}^{\mathrm{ss}}_{\mathrm{tot}}$ (in logscale)in the parameter space $(r,I)$, where distinct regions are seen to have different values of $\dot{S}^{\mathrm{ss}}_{\mathrm{tot}}$. The left panel (A2) shows $\dot{S}^{\mathrm{ss}}_{\mathrm{tot}}$ averaged over its $r$ values at a given $I$ versus the specified $I$, whereas the top panel (A3) depicts $\dot{S}^{\mathrm{ss}}_{\mathrm{tot}}$ averaged over its $I$ values at a given $r$ versus the specified $r$. (B1) The heat map of $\dot{S}^{\mathrm{ss}}_{\mathrm{tot}}$ in the smaller parameter space $(r,I)$ (marked by a rectangular box in A1) where stochastic complex oscillations are observed. As depicted by the time series $x(t)$ in $a$--$f$ (see Fig.~\ref{hrm1}~B) and $g$--$k$, $\dot{S}^{\mathrm{ss}}_{\mathrm{tot}}$ increases with more complex bursting of spikes. The left panel (B2) shows $\dot{S}^{\mathrm{ss}}_{\mathrm{tot}}$ averaged over its $r$ values at a given $I$ versus the specified $I$, whereas the top panel (B3) illustrates $\dot{S}^{\mathrm{ss}}_{\mathrm{tot}}$ averaged over its $I$ values at a given $r$ versus the specified $r$. In panel (B3), we fit the decreasing trend in the log-log plot of $\dot{S}^{\mathrm{ss}}_{\mathrm{tot}}$ versus $r$ with a power-law relation $\dot{S}^{\mathrm{ss}}_{\mathrm{tot}}(r)\sim r^{-\gamma}$, with power-law exponent $\gamma$. We find $\gamma=0.46$ with the goodness-of-fit $r^2$ score equals 0.81. We observe an order-to-disorder transition from organized, periodic bursts of spikes to unorganized chaotic (aperiodic) oscillations with a transition point around $r\approx 0.002$.  }}
    \label{hrm2}
\end{figure*}


In Fig.~\ref{hrm1}(A), we present a heat map of $\mathcal{W}(r,I)$ on a $220\times 260$ grid, covering the parameter range $I\in [1.0,8.0]$ and $r \in [0.0001,0.05]$. $\mathcal{W}$ is found to have a significant dependence on the values of $I$ and $r$, indicating regions of varying inter-spike interval reflective of variations in the periodicity of $x(t)$ across $(r,I)$. That is, distinct regions colored differently suggestive of diverse patterns of oscillations are evident. Our stochastic analysis is informed by previous findings (deterministic case) reported in Ref.~\cite{gonzalez2007complex}. On comparison of the bifurcation diagram of Fig.~5 in Ref.~\cite{gonzalez2007complex} with our Fig.~\ref{hrm1}(A), the region below the horizontal line around $I\approx 1.2$ corresponds to region A, the region bounded by lines around $I\approx 1.2$ and $I\approx 4.5$ to B, the region bounded by lines around $I\approx 4.5$ and $I\approx 5.0$ to C and above $I\approx 5.0$ to region D. These regions are characterized by distinct stability criteria. The fixed points in regions A and C are stable equilibrium, whereas those in B and D are unstable equilibrium~\cite{gonzalez2007complex}. Eigenvalue analysis reveals further details on the nature of the dynamics around the fixed points; stable region A consisting of equilibrium which are sink and spiral sinks, unstable region B of saddle and a spiral source, stable region C of spiral sinks and region D of spiral source~\cite{gonzalez2007complex}. We refer the readers to Ref.~\cite{gonzalez2007complex} for a detailed eigenvalue analysis of the systems' Jacobian.

Notably, interesting features are seen in the smaller parameter domain of $I\in [1.2,4.5]$ and $r \in [0.0001,0.01]$. This domain corresponds to the unstable region B, where complex nonlinear bifurcation structures emerge, as reported in Ref.~\cite{gonzalez2007complex}. The heat map of $\mathcal{W}$ in Fig.~\ref{hrm1}(B) offers further insights into this interesting domain (marked by a rectangular box in Fig.~\ref{hrm1}(A)). We illustrate representative stochastic trajectories $x(t)$ for different values of $I$ at a specified $r=0.002$, denoted by markers $a$ through $f$. These trajectories depict periodic firings characterized by bursts of spikes interspersed with lapses of quiescence. Such bursting patterns, observed in trajectories $x(t)$ from $a$--$f$, are known as block-structured dynamics~\cite{gonzalez2005block}. This block-structured dynamics signifies a complex bifurcation structure, where groups or blocks of similar periodicity oscillations emerge. At a given $r$, the dynamics of $x(t)$ in a segment of $I$ exhibit the same periodicity (indicated by specific colors), contributing to the block-structured dynamics. It is evident from the trajectories $x(t)$ that $\mathcal{W}$ increases as the number of spikes or periodicity decreases. We note that $x(t)$ in $d$--$f$ shows the same red color in the heat map due to the limited available range in the colormap. Ideally, $\mathcal{W}$ will be the largest in the case of simple periodic firing characterized by single spikes. Thus, $\mathcal{W}$ characterizes the dynamical behavior of $x(t)$ of the sHR oscillator model~\eqref{eq:shrm}. 

We now proceed to analyze the interplay of effects arising due to $(r,I)$ on the steady-state total entropy production rate ($\dot{S}^{\mathrm{ss}}_{\mathrm{tot}}$). The panels (A1) and (B1) of Figs.~\ref{hrm2}(A)$\&$(B) show $\dot{S}^{\mathrm{ss}}_{\mathrm{tot}}$ across the two-dimensional $r-I$ plane considered in Figs.~\ref{hrm1}(A)$\&$(B), respectively. The color bar indicates the magnitude of $\dot{S}^{\mathrm{ss}}_{\mathrm{tot}}$ in log scale with base 10. 
As depicted in the panel (A1), the patterns in $\dot{S}^{\mathrm{ss}}_{\mathrm{tot}}$ reflect the various dynamical regimes of the sHR oscillator as characterized by the function $\mathcal{W}$ discussed above.  
This is confirmed by the left panel (A2) of Fig.~\ref{hrm2}(A) showing $\dot{S}^{\mathrm{ss}}_{\mathrm{tot}}$ (in logscale) versus $I$. This plot is obtained by averaging $\dot{S}^{\mathrm{ss}}_{\mathrm{tot}}$ values for all $r$ at the given $I$. We see that $\dot{S}^{\mathrm{ss}}_{\mathrm{tot}}$ has a non-monotonic relation with $I$. The values of $\dot{S}^{\mathrm{ss}}_{\mathrm{tot}}$ oscillates with $I$. As $I$ varies, we clearly observe distinct inflection points that divide the $r-I$ region into distinct dynamical regimes, as discussed above.
Similarly, in the top panel (A3) of Fig.~\ref{hrm2}(A), we plot $\dot{S}^{\mathrm{ss}}_{\mathrm{tot}}$ versus $r$ (loglog scale) by averaging $\dot{S}^{\mathrm{ss}}_{\mathrm{tot}}$ values for all $I$ at a given $r$. $\dot{S}^{\mathrm{ss}}_{\mathrm{tot}}$ seems to follow a linearly decreasing trend till $r\approx 0.002$,  beyond which $\dot{S}^{\mathrm{ss}}_{\mathrm{tot}}$ shows non decreasing behavior.



From the panel (A1) of Fig.~\ref{hrm2}(A), in the region $1.2<I<4.0$, wee see complex patches of lower and higher values of $\dot{S}^{\mathrm{ss}}_{\mathrm{tot}}$. This region encompasses cycles and chaotic dynamics interwoven in complex structures, as previously characterized in Ref.~\cite{gonzalez2007complex} and by our results shown in Fig.~\ref{hrm1}(B). In the panel (B1) of Fig.~\ref{hrm2}(B), we take a closer look at the behavior of $\dot{S}^{\mathrm{ss}}_{\mathrm{tot}}$ within the smaller parameter space (marked by a rectangular box in panel (A1) of Fig.~\ref{hrm2}(A)). This smaller parameter space is the same as considered in Fig.~\ref{hrm1}(B). Intriguing patterns of $\dot{S}^{\mathrm{ss}}_{\mathrm{tot}}$ emerge, which can be associated with our previous observations of complex oscillations characterized by bursts of multiple spikes in Fig.~\ref{hrm1}(B). The markers $a$--$f$ denote the same block-structured dynamics discussed in Fig.~\ref{hrm1}(B). In the left panel (B2) of Fig.~\ref{hrm2}(B), we plot $\dot{S}^{\mathrm{ss}}_{\mathrm{tot}}$ (in logscale) versus $I$ by averaging its values for all $r$ at a given $I$.  A sharp transition exists around $I\approx 1.3$, which marks the transition from stable region A to unstable region B~\cite{gonzalez2007complex}. $\dot{S}^{\mathrm{ss}}_{\mathrm{tot}}$ gradually increases till $I\approx 2.6$, beyond which it plateaus and decreases. Examining the trajectories $x(t)$ from markers $d$--$f$, $\dot{S}^{\mathrm{ss}}_{\mathrm{tot}}$ generally increases with bursts of more spikes at a given $r$. In neurons, the block structures serve as fundamental encoders of messages within the stochastic dynamics of the membrane potential $x(t)$ traversing along the axon~\cite{gonzalez2007complex}. Since $\dot{S}^{\mathrm{ss}}_{\mathrm{tot}}$ rises with increasingly complex bursting dynamics, our results suggest that encoding more information by neurons entails a higher thermodynamic cost. The plateau region and decreasing tail in  $\dot{S}^{\mathrm{ss}}_{\mathrm{tot}}$ beyond $I\approx 2.6$ arise due to additional contributions from chaos, which are aperiodic oscillations.

To see chaotic oscillations, we plot stochastic trajectories $x(t)$ for various values of $r$ at a fixed $I=3.25$, depicted in markers $g$--$k$. A similar increasing trend in $\dot{S}^{\mathrm{ss}}_{\mathrm{tot}}$ with bursts containing more spikes is observed. Notably, a transition from chaotic to bursting dynamics is apparent in the trajectories $x(t)$ from $i$ and $h$. Chaos is evident in $x(t)$ of the markers $i,j$, and $k$ beyond $r\approx 0.002$.
The top panel (B3) of Fig.~\ref{hrm2}(B) confirm these two observations. Moreover, the decreasing trend in the log-log plot of $\dot{S}^{\mathrm{ss}}_{\mathrm{tot}}$ versus $r$ can be described by a power-law relation of $\dot{S}^{\mathrm{ss}}_{\mathrm{tot}}(r)\sim r^{-\gamma}$, where $\gamma$ is the power-law exponent. We fit the data of $\dot{S}^{\mathrm{ss}}_{\mathrm{tot}}$ with such a power-law, resulting in $\gamma=0.46$ (see the solid blue line for fitting in panel (B3). We measure the goodness-of-fit with $r^2$ score, which is equal to 0.81 for our fit. From the panel (B3), the two distinct behavior of $\dot{S}^{\mathrm{ss}}_{\mathrm{tot}}$ indicates an order-to-disorder transition from periodic (organized) bursts of spikes to chaotic oscillations (unorganized regime) with a transition point around $r\approx 0.002$. Our results show a higher thermodynamic cost for the regular or ordered periodic bursting behavior than irregular or disordered chaotic dynamics. This again confirms our proposition that complex-ordered oscillations involve more dissipation.
As previously shown in Ref.~\cite{gonzalez2007complex}, the inter-spike interval $\mathcal{T}$ has a more complex pattern with respect to $r$ than with $I$ in the smaller parameter domain considered. Our estimated $\dot{S}^{\mathrm{ss}}_{\mathrm{tot}}$ also confirms this. 

To summarize, our analysis of steady-state total entropy production rate ($\dot{S}^{\mathrm{ss}}_{\mathrm{tot}}$) in the nonequilibrium models of the intracellular Ca$^{2+}$ oscillation (ICO) model~\eqref{eq:camodel} and the scaled Hindmarsh-Rose (sHR) oscillator model~\eqref{eq:shrm} reveals that complex oscillations entail higher values of $\dot{S}^{\mathrm{ss}}_{\mathrm{tot}}$ than simple periodic oscillations, implying that complex oscillations incur higher dissipation (or energy consumption). 
\section{Concluding remarks}
\label{sec:conclu}
Nonequilibrium thermodynamics provides a powerful framework for understanding thermodynamic properties of systems far from equilibrium~\cite{fang2019nonequilibrium, qian2005thermodynamics}, prevalent in various natural systems including biological, chemical, and physical systems. These systems often exhibit fluctuations-driven complex patterns or structures in space and/or time, known as dissipative structures. These dissipative structures are sustained through the dissipation of energy
~\cite{prigogine1977self,haken1987synergetics}. The quantification of energy dissipation in nonequilibrium systems is often achieved through the measurement of entropy production. The investigation of entropy production in nonlinear, nonequilibrium systems constitutes a vibrant field of research, incorporating both theoretical and experimental studies. Previous studies have estimated entropy
production in nonlinear systems like cyclic population
models~\cite{andrae2010entropy}, the Schl$\mathrm{\ddot{o}}$gl model~\cite{chen2013entropy} describing nonequilibrium phase transition, and chaotic chemical systems~\cite{li2012potential,gaspard2020stochastic,zhang2020dynamic}. While dissipation in stochastic simple periodic oscillations has been well-explored, the study of dissipation in stochastic complex oscillations, such as bursting and spiking dynamics, remains relatively sparse. In this paper, we aim to fill this research gap by investigating dissipation in two nonlinear, nonequilibrium models of biological systems with complex oscillatory dynamics: intracellular calcium (Ca$^{2+}$) oscillation and Hindmarsh-Rose (sHR) neuron dynamics models.

By formulating open chemical reaction systems for the two nonlinear models using stochastic modeling via the chemical Langevin equation, we have quantified the steady-state total entropy production rate ($\dot{S}^{\mathrm{ss}}_{\mathrm{tot}}$) in these two nonequilibrium systems. $\dot{S}^{\mathrm{ss}}_{\mathrm{tot}}$ is found to have a significant dependence on the system parameter values considered, indicating its dependence on diverse dynamical structures. Importantly, our findings reveal that complex oscillations, such as bursting and quasi-periodic oscillations, entail higher values of $\dot{S}^{\mathrm{ss}}_{\mathrm{tot}}$ than simple periodic oscillations, suggesting that complex oscillations incur a higher thermodynamic cost. Notably, in the scaled Hindmarsh-Rose (sHR) oscillator model, we observe an order-to-disorder transition from periodic (organized) bursts of spikes to chaotic (unorganized) oscillations with distinct behaviors of $\dot{S}^{\mathrm{ss}}_{\mathrm{tot}}$. Our results show that $\dot{S}^{\mathrm{ss}}_{\mathrm{tot}}$ of periodic bursts of spikes is larger than that of chaotic oscillations, implying that the nonequilibrium system of sHR model dissipates more energy to maintain the organized behavior of bursts than the irregular chaotic oscillations.    
While the number of spikes in the bursting dynamics simply increases in the sHR oscillator model, the nature of the dynamical structure changes with system parameters in the case of the ICO model, namely, quasi-periodicity with multiple frequencies and bursting with a sharp spike followed by small spikes. The two different nonequilibrium models could account for the generality of our finding on higher entropy production rates in the context of nonlinear complex oscillations. While increased bursting of spikes implies rich information coding by neurons, complex intracellular Ca$^{2+}$ oscillations indicate complex cellular regulatory or signaling processes by Ca$^{2+}$. Hence, our results of complex oscillations having higher values of $\dot{S}^{\mathrm{ss}}_{\mathrm{tot}}$ reflect that the complex processes of neuronal cells or intracellular Ca$^{2+}$ involve a higher thermodynamic cost. Our findings contribute to a deeper understanding of energy dissipation in nonlinear, nonequilibrium biological systems with complex oscillatory dynamics. It will be interesting to confirm our theoretical results of higher entropy production rates in complex oscillations for real experimental data of neuronal bursts or intracellular Ca$^{2+}$ oscillation.

In the present work, we have not investigated the nonequilibrium dynamical origins of stochastic nonlinear oscillators with complex oscillations. One may address this with a theoretical framework such as potential and flux landscape~\cite{wang2015landscape}. In nonequilibrium open systems, potential landscape and curl probability flux are found significant to describe their underlying global dynamics~\cite{wang2015landscape}. It is intriguing to investigate the interplay of potential landscape and curl flux in the nonequilibrium CRN with complex oscillations. Further investigation in this direction will be addressed in future work. Another future outlook of the present paper could be to explore the link between nonequilibrium thermodynamics and complexity theory. Entropy production rate has recently been used to evaluate the complexity and robustness of cancer~\cite{nieto2023rate}. The link between entropy production rate and complexity remains not very clear, necessitating further research.   

\section*{Authors' contributions}
{\noindent}ALC-Conceptualization; formal investigation (lead); writing-original draft preparation; writing—review and editing (equal). PM-formal investigation (support); writing—review and editing (equal). SK-formal investigation (support); writing—review and editing (equal). RKBS-writing—review and editing (equal).
\section*{Acknowledgments}
\noindent This research is supported by an appointment of ALC to the YST program at the APCTP through the Science and Technology Promotion Fund and Lottery Fund of the Korean Government and the Korean Local governments-Gyeongsangbuk-do Province and Pohang City. PM acknowledges financial support from JNU and UGC (New Delhi, India). SK acknowledges JNU and ICMR (New Delhi, India) for a Senior Research Fellowship vide award number 3/1/3/9/M/2022-NCD-II.
\section*{Conflict of Interests }
\noindent The authors have no conflicts to disclose.
\section*{Data Availability Statement}
\noindent The data that support the findings of this study are available within the article itself.
\appendix

\section{}
\label{sec:appen0}
For a chemical reaction network~\eqref{crn1} described by the state vector $\mathbf{X}=\{X_1,X_2,\dots,X_k\}$ of species populations, the chemical Master equation (CME) models the dynamics of $\mathbf{X}$ as a Markov process in discrete states and continuous time. Suppose $\mathcal{P}(\mathbf{X},t)$ denotes the probability of finding the chemical reaction system in configuration $\mathbf{X}$ at any instant of time $t$. The CME describing the time evolution of $\mathcal{P}(\mathbf{X},t)$ is given by
\begin{equation}
    \frac{\partial \mathcal{P}(\mathbf{X},t)}{\partial t}=\Omega \sum_{m}\left(\prod_{i=1}^k \mathbb{E}_{i}^{-S_{ij}}-1\right)\nu_m(\mathbf{X}) \mathcal{P}(\mathbf{X},t). \label{eq:meqn}
\end{equation}
Here, $\Omega$ represents the system size, $\mathbb{E}$ is the step operator defined as $\mathbb{E}_i^{\gamma} \ g(X_1,X_2,\dots,X_i,\dots)=g(X_1,X_2,\dots,X_i+\gamma,\dots)$ for integer $\gamma$ \cite{van1992stochastic}, $S_{ij}$ denotes the stoichiometric matrix, and $\nu_m(\mathbf{X})$ is the propensity function of reaction step $m$ defined as:
\begin{equation}
    \nu_m(\mathbf{X})=\eta_m \prod_{l\in \textbf{X}} \frac{1}{\Omega ^{\sigma^{X}_{lm}}} \frac{X_l!}{(X_l-\sigma^{X}_{lm})!}, \label{eq:pro}
\end{equation} 
with reaction rates $\eta_m$ and stoichiometric constant $\sigma^{X}_{lm}$. Solving a multivariate CME analytically is generally infeasible. To simplify the CME, one can treat the species $X_i$ as real continuous variables defined by $x_i=X_i/\Omega$ (continuous state space). By employing system-size expansion~\cite{van1992stochastic}, one can derive the Fokker-Planck (FP) equation from the CME~\eqref{eq:meqn}, obtaining drift and diffusion terms. Alternatively, one can use the Kramers-Moyal expansion and retain the lowest-order terms \cite{risken1984fokker} to get the FP equation. 
\section{}
\label{app:A1}
\renewcommand{\thefigure}{C.\arabic{figure}}
\setcounter{figure}{0}
\begin{table*}
\centering
\caption{Differential equations of the intracellular calcium oscillation model~\eqref{eq:camodel} with the corresponding set of reaction channels and propensity functions.} 
\label{table:two}
\begin{tabular}{|p{6cm}|p{8cm}|p{3cm}|}
\hline
Ordinary differential equations & Reaction channels  & Propensity functions  \\
\hline
\small{$\frac{dx}{dt}=V_0+V_1\beta-V_2+V_3+k_fy-kx $} & &  \\
&\small{\ce{A -> X }} & $V_0$ \\
& & \\
&\small{\ce{B -> X }} & $V_1\beta $  \\
& & \\
&\small{\ce{M +X -> M^* + X} (slow), \ce{M^* +X -> M + C} (fast)} & $ V_2$ \\
& & \\
&\small{\ce{D -> X }}& $V_3$ \\
& & \\
$\frac{dy}{dt}=V_2-V_3-k_fy$ & &\\
&\small{\ce{Y -> X }} & $k_f y $\\& & \\
&\small{\ce{X -> E }} & $kx$\\
& & \\
&\small{\ce{F -> Y }} & $V_2$\\
& & \\
&\small{\ce{M + Y -> M^* + Y} (slow), \ce{M^* +Y -> M + G} (fast)}  & $V_3$\\
& & \\

&\small{\ce{Y -> H}} & $k_fy$\\
& & \\

\small{$\frac{dz}{dt}=\beta V_4-V_5-\epsilon z$} & & \\ 
&\small{\ce{I -> Z }} &$\beta V_4$\\
& & \\
&\small{\ce{M + Z -> M^* + Z} (slow), \ce{M^* +Z -> M + J} (fast)} &$V_5$\\
& & \\
&\small{\ce{Z -> K }} &$\epsilon z$\\
& & \\
\hline
\end{tabular}
\end{table*}
\begin{table*}
\centering
\label{table:one}
\caption{Differential equations of the scaled Hindmarsh-Rose oscillator model~\eqref{eq:shrm} with the corresponding set of reaction channels and propensity functions.} 
\label{table:one}
\begin{tabular}{|p{5.8cm}|p{8.7cm}|p{3cm}|}
\hline
Ordinary differential equations & Reaction channels  & Propensity functions  \\
\hline
\small{$\frac{dx}{dt}=(y-\alpha_y)+3(x-\alpha_x)^{2}-(x-\alpha_x)^{3}-z+I $} & &  \\
&\small{\ce{D_1 +Y -> D_1^* +Y (slow), D_1^* +A -> X +D_1 (fast)}}& y \\
& & \\
&\small{\ce{3X -> R }} & $\frac{1}{3}x(x-\frac{1}{\Omega})(x-\frac{2}{\Omega})$  \\
& & \\
&\small{\ce{2X -> 3X }} & $ 3(\alpha_{x}+1)x(x-\frac{1}{\Omega})$ \\
& & \\
&\small{\ce{X -> R }}& $3\alpha_{x}(2+\alpha_{x})x $ \\
& & \\
&\small{\ce{D_2 +Z -> D_2^* + Z} (slow), \ce{D_2^* +X -> D_2 + R} (fast)} & z \\
& & \\
&\small{\ce{A -> X }} &$\alpha_{x}^{3}+3\alpha_{x}^{2}+I$ \\
& & \\
&\small{\ce{A + D_3 -> D_3^* + A} (slow), \ce{D_3^* +X -> D_3 + R} (fast)}  & $\alpha_{y}$ \\
& & \\

$\frac{dy}{dt}=1-5(x-\alpha_x)^{2}-(y-\alpha_y)$ & &\\
&\small{\ce{B -> Y }} & $1 +\alpha_y $\\
& & \\
&\small{\ce{2X + D_4 -> D_4^* + 2X} (slow), \ce{D_4^* +Y -> D_4 + R} (fast)}   & $5x(x-\frac{1}{\Omega})$\\
& & \\
&\small{\ce{D_5 +X -> D_5^* +X (slow), D_5^* +B -> Y +D_5} (fast)} & $10\alpha_{x}x$\\
& & \\
&\small{\ce{Y -> R }} & $y$\\
& & \\
&\small{\ce{B + D_6 -> D_6^* + B} (slow), \ce{D_6^* +Y -> D_6 + R} (fast)} & $5\alpha_{x}^{2}$\\
& & \\
 
\small{$\frac{dz}{dt}=r[4(x-\alpha_x+\frac{8}{5})-z]$} & & \\ 
&\small{\ce{D_7 +X -> D_7^* +X (slow), D_7^* +C -> Z +D_7  } (fast)} &$4rx$\\
& & \\
&\small{\ce{Z -> C }} &$rz$\\
& & \\
&\small{\ce{C + D_8 -> D_8^* + C} (slow), \ce{D_8^* +Z -> R + D_8} (fast)} &$4r\alpha_x$\\
& & \\
&\small{\ce{C -> Z }} &$\frac{32r}{5}$\\
& & \\
\hline
\end{tabular}
\end{table*}

 \textbf{Intracellular calcium (Ca$^{2+}$) oscillation (ICO) model:} In the nonlinear ODEs~\eqref{eq:camodel} describing the ICO model, the dynamical equations  $\frac{dx}{dt}$ and $\frac{dy}{dt}$ represent the rate of change of the concentrations of cytosolic Ca$^{2+}$ ($x(t)$) and Ca$^{2+}$ stored in the internal pool ($y(t)$), respectively. We present a breakdown of the terms involved as follows~\cite{chanu2024exploring}: $V_0$ denotes the constant Ca$^{2+}$ supply from the extracellular medium. The parameter $\beta$ represents the degree of cell stimulation by an agonist (e.g., hormone or neurotransmitter). $V_1$ represents the maximum rate of stimulus-activated Ca$^{2+}$ entry from the extracellular medium. The rate $V_2 \ (V_3)$ corresponds to Ca$^{2+}$ pumping from the cytosol into the internal pool (release of Ca$^{2+}$ from the internal pool to the cytosol). $V_{M2}$ and $V_{M3}$ denote their maximum values. Parameters $k_2, \  k_y, \ k_x$, and $k_z$ represent the threshold values for pumping, release, and activation of release by Ca$^{2+}$ and InsP$_3$, respectively. The rate constant $k_f$ measures the passive, linear leak of $y$ into $x$, and $k$ signifies the linear transport of cytosolic Ca$^{2+}$ into the extracellular medium. The evolution equation $\frac{dz}{dt}$ in the ODEs~\eqref{eq:camodel} represents the rate of change of InsP$_3$. $V_4$ denotes the maximum rate of stimulus-induced InsP$_3$ synthesis, and $V_5$ represents the phosphorylation rate of InsP$_3$ by the 3-kinase, an InsP$_3$ metabolising enzyme. The decrease of InsP$_3$ is driven by its hydrolysis by calcium-dependent 3-kinase. $k_5$ denotes the half-saturation constant. Stimulation of InsP$_3$ 3-kinase activity (through a Ca$^{2+}$/calmodulin complex) is represented by a Hill-form term with $k_d$ as the threshold level of Ca$^{2+}$. The term $-\epsilon Z$ accounts for the metabolism of InsP$_3$ by 5-phosphatase, independent of Ca$^{2+}$. Additionally, cooperative processes in Ca$^{2+}$ release from internal stores into the cytosol and phosphorylation of InsP$_3$ by 3-kinase are reflected in $V_3$ and $V_5$, incorporating Hill-coefficients $m$, $n$ and $p$. 
 
The deterministic drift $\textbf{f}$ for the ICO model~\eqref{eq:camodel} is
\begin{equation}
    \textbf{f}=\left(
    \begin{array}{c}
    V_0+V_1\beta-V_2+V_3+k_fy-kx\\
    V_2-V_3-k_fy\\
    \beta V_4-V_5-\epsilon z 
\end{array}
\right).
\end{equation}
Using Eqs.~\eqref{eq:d}$\&$\eqref{eq:alpha}, the diffusion matrix $\textbf{D}$ and the term $\alpha$ are
\tiny
\begin{equation} 
\textbf{D}=\left(
\begin{array}{ccc}
 V_0+V_1\beta+V_2+V_3+k_fy+kx & 0 & 0 \\
 0 & V_2+V_3+k_fy & 0 \\
 0 & 0 & \beta V_4+V_5+\epsilon z \\
\end{array}
\right),
\end{equation}
\normalsize
\begin{equation}
    \alpha=\left(
    \begin{array}{c}
    V_0+V_1\beta-V_2+V_3+k_fy-kx\\
    V_2-V_3-k_fy\\
    \beta V_4-V_5-\epsilon z 
\end{array}
\right)-\frac{1}{2\Omega}\left(
    \begin{array}{c}
    k+k_f\\
    k_f\\
    \epsilon  
\end{array} \right).
\end{equation}\\
{\noindent}\textbf{Hindmarsh-Rose (HR) neuron model:} In the neuronal cells of the pond snail \textit{Lymnaea}, experimental observations revealed that a short electric pulse could depolarize a quiescent neuron, initiating an action potential and resulting in a burst that persisted longer than the initial stimulus~\cite{hindmarsh1984model}. The HR model~\cite{hindmarsh1982model} was developed to elucidate such complex bursting dynamics observed in neurons. It is a three-dimensional model described by the following set of coupled nonlinear ODEs:
\begin{align}
&\dot{x}=y+3x^{2}-x^{3}-z+I, \nonumber \\
& \dot{y}=1-5x^{2}-y, \label{eq:hrm}\\
&\dot{z}=r\left[4\left(x+\frac{8}{5}\right)-z\right], \nonumber
\end{align}
where the dynamical variables $x,~y$, and $z$ represent the membrane potential, ion transports through fast and slow ion channels, respectively. Hindmarsh and Rose modified the original Fitzhugh's Bonhoeffer-van der Pol model~\cite{hindmarsh1984model,fitzhugh1961impulses,nagumo1962active} by replacing a linear term with a parabolic term in the recovery variable, denoted by $y(t)$ (see the second equation of~\eqref{eq:hrm}). Additionally, to account for observations like hyper-polarization-induced cessation of bursting and burst discharges triggered by depolarizing current pulses, an extra differential equation with the adaptation current, denoted by $z(t)$, was introduced (see the third equation of~\eqref{eq:hrm}). Depending on the parameters $I$ and $r$, the HR model~\eqref{eq:hrm} exhibits spiking patterns, either single spikes or bursts of periodic spikes with periodicity number $p$~\cite{gonzalez2007complex}. The complex bursting patterns, including chaotic behavior, correspond to diverse neuro-physiological states such as rapid responses of the nervous system by switching between different dynamical states.

For the scaled Hindmarsh-Rose (sHR) oscillator model~\eqref{eq:shrm}, we calculate the deterministic drift as 
\begin{widetext}
    \small
$\textbf{f}=\left(
\begin{array}{c}
 I+\alpha_{x}^3+3 \alpha_{x}^2+3 (\alpha_{x}+1) x \left(x-\frac{1}{\Omega}\right)-3 (\alpha_{x}+2) \alpha_{x} x-\alpha_{y}- x \left(x-\frac{2}{\Omega}\right) \left(x-\frac{1}{\Omega}\right)+y-z \\
 -5 \alpha_{x}^2+10 \alpha_{x} x+\alpha_{y}-5 x \left(x-\frac{1}{\Omega}\right)-y+1 \\
 -4 \alpha_{x} r+4 r x-r z+\frac{32 r}{5} \\
\end{array}
\right).$
\end{widetext}
Using Eqs.~\eqref{eq:d}$\&$\eqref{eq:alpha}, the diffusion matrix $\textbf{D}$ and the term $\alpha$ are calculated as
\begin{widetext}
    \tiny 
$ \textbf{D}=\left(
\begin{array}{ccc}
 I+\alpha_{x}^3+3 \alpha_{x}^2+3 (\alpha_{x}+1) x \left(x-\frac{1}{\Omega}\right)+3 (\alpha_{x}+2) \alpha_{x} x+\alpha_{y}+3 x \left(x-\frac{1}{\Omega}\right) \left(x-\frac{2}{\Omega}\right)+y+z & 0 & 0 \\
 0 & 5 \alpha_{x}^2+10 \alpha_{x} x+\alpha_{y}+5 x \left(x-\frac{1}{\Omega}\right)+y+1 & 0 \\
 0 & 0 & 4 \alpha_{x} r+4 r x+r z+\frac{32 r}{5} \\
\end{array}
\right),
$
\end{widetext}

\begin{widetext}
$\alpha = \textbf{f}-\frac{1}{2\Omega} \left(
\begin{array}{c}
 3 (\alpha_x+1) \left(x-\frac{1}{\Omega}\right)+3 (\alpha_x+1) x+3 \alpha_x (\alpha_x+2)+3 x \left(x-\frac{1}{\Omega}\right)+3 x \left(x-\frac{2}{\Omega}\right)+3 \left(x-\frac{1}{\Omega}\right) \left(x-\frac{2}{\Omega}\right) \\
 1 \\
 r \\
\end{array}
\right) $
\end{widetext}

\section{Stochastic Modeling with Chemical Langevin Equation (CLE)}
\label{app:Acle}
Consider a well-stirred chemically reacting system of size $\Omega$ containing three molecular species of populations $X$, $Y$, and $Z$, where the state vector of their populations in the system is $\textbf{X}=[X(t), Y(t), Z(t)]^\mathrm{T}$. Here, T denotes transpose. Suppose $X$, $Y$, and $Z$ species interact through a set of $M$ chemical reactions as: $\ce{$l_{1j}X$ + $m_{1j}Y$ + $n_{1j}Z$ ->[$k_j$] $l_{2j}X$ + $m_{2j}Y$ + $n_{2j}Z$}$, where $k_j\ ;\ j=1,2,\dots,M$ denotes the classical rate constant of the $j^{\mathrm{th}}$ reaction. The sets $\{l_1,m_1,n_1\}$ and $\{l_2,m_2,n_2\}$ represent the numbers of reactant and product molecules, respectively. The probability of a reaction $j$ occurring within $\Omega$ in the next infinitesimal time interval $(t,t+dt)$ is given by $a_j[\textbf{X}]dt$, where $a_j[\textbf{X}]$ is the propensity function for reaction $j$.

The chemical Langevin equation (CLE) is a stochastic differential equation that, in general, has the form~\cite{gillespie2000chemical}:
\begin{equation}
    \frac{d X_i(t)}{dt}=\sum_{j=1}^{M} \nu_{ji} \ a_j[\textbf{X}]+\sum_{j=1}^{M}\nu_{ji}\ a_j^{1/2}[\textbf{X}]\ \xi_j(t), \label{eq:cl}
\end{equation}
where $\nu_{ji}$ denotes the stoichiometric coefficient for species $X_i$ in reaction $j$, and $\xi_j$(t) are random noise parameters with $\xi_j(t)=\displaystyle \lim_{dt\rightarrow 0} \mathcal{N}(0,1/dt)$~\cite{gillespie2000chemical,gillespie1996multivariate}. On the right-hand side of Eq.~\eqref{eq:cl}, the first term accounts for the deterministic part, whereas the second term represents the stochastic term. $X_i$ now becomes continuous variables.

\section{Methods}
\label{app:method}
\begin{algorithm}[H]
	\caption{Estimating steady-state total entropy production rate ($\dot{S}^{\mathrm{ss}}_{\mathrm{tot}}$)\\}
	\begin{algorithmic}[1]
		
		\State $\mathbf{x_0} \leftarrow$ Initial state vector
		\State $\mathbf{x_t} \leftarrow$ Trajectory vector 
		\State $\alpha \leftarrow$ Force Vector
		\State D$ \leftarrow$ Diffusion Matrix

		\Function{SolveChemicalLangevin}{$\mathbf{x}_0$, $0:t$, (params(r,I), $\Omega$)}
		\State model $\leftarrow$ $\mathbf{F(x_t)}$ + $\mathbf{G(x_t)\mathbf{dW_t}}$
		\State $\mathbf{x_t}$ $\leftarrow$ \Call{solve\_ivp}{\text{model}} \Comment{trajectory vector $\mathbf{x_t}$}
		\State \textbf{return} $\mathbf{x_t}$
		\EndFunction

		\Function{KernelDensityEstimator}{$\mathbf{x_t}$}
		\State data $\leftarrow$ $\mathbf{x_t}[\text{startindex, endindex}]$ \Comment{Segment of trajectory for density calculation}
		\State kde $\leftarrow$ \Call{KernelDensity}{\text{bandwidth, kernel=`epanechnikov'}} \Comment{Python package for Multi-variate KernelDensity}
		\State kde.fit($\mathbf{x_t}$)
		\State density $\leftarrow$ kde.score\_samples(data)
		\State \textbf{return} $\mathbf{density}$
		\EndFunction
		
		\Function{main()}{}
		\State $x_t$ $\leftarrow$ \Call{SolveChemicalLangevin}{$\textbf{x}_0$, $0:t$, (params(r,I), $\Omega$)}
		\State density $\leftarrow$ \Call{KernelDensityEstimator}{$\mathbf{x_t}$}
		\State Using $\alpha, D$ $\&$ density, estimate $\hat{J}^{\mathrm{ss}},\dot{S}^{\mathrm{ss}}_{\mathrm{tot}}$         
		\State \textbf{return} $\dot{S}^{\mathrm{ss}}_{\mathrm{tot}}$
		\EndFunction
	\end{algorithmic} 
\end{algorithm}
\textbf{Algorithm 1} presents a pseudo-code detailing the steps we have followed to estimate the steady-state total entropy production rate ($\dot{S}^{\mathrm{ss}}_{\mathrm{tot}}$) for the nonequilibrium systems of intracellular calcium oscillation (ICO) model~\eqref{eq:camodel} and the scaled Hindmarsh-Rose (sHR) oscillator model~\eqref{eq:shrm}. 
\renewcommand{\thefigure}{D.\arabic{figure}}
\setcounter{figure}{0}

To obtain the trajectories $x(t)$, we numerically integrate the CLEs of Eqs.~\eqref{eq:cle22}$\&$\eqref{eq:cle2} using the standard module \texttt{solve\_ivp} from the \texttt{SciPy} package in Python 3, which uses, by default, an explicit Runge-Kutta integration method of order 5~\cite{2020SciPy-NMeth}. For the ICO model~\eqref{eq:camodel}, we take initial values $x_0=y_0=z_0 =0.5.$ In the case of the sHR oscillator model~\eqref{eq:shrm}, we choose the scaling parameters $\alpha_x =13$ and $\alpha_y =60.$ The initial vector $\mathbf{x_0}$ is: $x_0 = 12,~y_0 = 60,~z_0 = 3$.  

We use the method developed in Ref.~\cite{li2019quantifying} to estimate the various quantities of our interest. In Ref.~\cite{li2019quantifying}, the estimator for multivariate probability density $\hat{P}^{\mathrm{ss}}(\textbf{x})$ from a single long trajectory was shown to depend on the trajectory length taken, achieving a convergence for greater length. Considering this, we choose a long trajectory of $5\times 10^5$ data points. To estimate the $\hat{P}^{\mathrm{ss}}(\textbf{x})$ from the previously generated trajectory $x(t)$, we first employ the standard Python package of \texttt{KernelDensity} from the \texttt{sklearn.neighbors} module \cite{scikit-learn}. We set the kernel argument to ``\texttt{epanechnikov kernel}" and calculate the bandwidth using the method suggested in Bowman and Azzalini (Ch.~2, Page no.~32)~\cite{bowman1997applied}. Using Eq.~\eqref{eq:j}, we then estimate the-steady state probability current $\hat{J}^{\mathrm{ss}}(\textbf{x})$. Finally, using Eq.~\eqref{eq:epr}, the steady-state total entropy production rate, $\dot{S}^{\mathrm{ss}}_{\mathrm{tot}}$ is determined. For all simulations, we take system size $\Omega = 10^8$, and the total simulation time of [0,15000] with 500,000 data points between them. After discarding the initial transients, we choose a trajectory segment of length 450,000 data points for KDE evaluation.  


\renewcommand{\thefigure}{A.\arabic{figure}}
\setcounter{figure}{0}

\bibliographystyle{refstyl}
\bibliography{references}

\end{document}